\documentclass[longauth]{aa}

\usepackage{graphicx}
\usepackage{lscape}
\usepackage{longtable}
\usepackage{soul}
\usepackage{lipsum}
\usepackage{hyperref}
\hypersetup{
    colorlinks=true,
    citecolor=blue,
    }
\usepackage{chngcntr}

\usepackage{txfonts}
\newcommand{\orcid}[1]{\unskip\protect\href{https://orcid.org/#1}{\protect\includegraphics[width=8pt,clip]{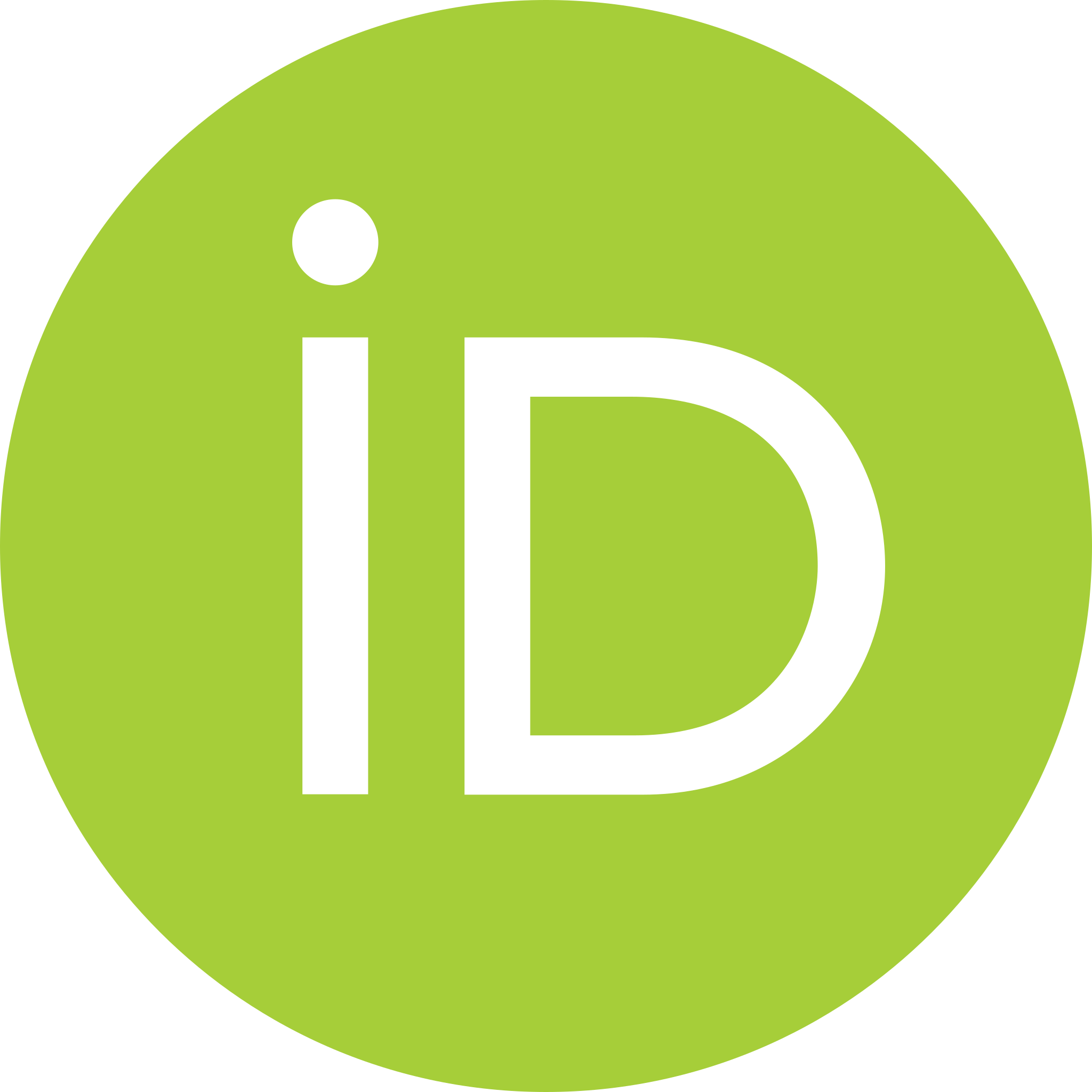}}}

\usepackage{float}
\usepackage{multirow} 

\newcommand{\Lsun}{$L_{\odot}$}
\newcommand{\Msun}{$M_{\odot}$}

\newcommand{\Mdot}{$\dot{M}$}
\newcommand{\Mdust}{${M_{\rm{dust}}}$}

\newcommand{\msunyr}{$\rm{M_{\sun} \, yr^{-1}}$}

\newcommand{\mic}{$\mu$m}

\newcommand{\mstar}{$M_{*}$}

\newcommand{\mdisk}{$M_{\rm{disk}}$}

\newcommand{\Lacc}{$L_{\rm{acc}}$}
\newcommand{\LaccL}{$L_{\rm{acc}}$/$L_{*}$}

\newcommand{\Lstar}{$L_{*}$}

\usepackage{xcolor}

\begin{document}

\title{MINDS. A transition from H$_2$O to C$_2$H$_2$ dominated spectra with decreasing stellar luminosity}
\titlerunning{}

\author{
Sierra L. Grant\orcid{0000-0002-4022-4899}\inst{1,2}  
\and Milou Temmink\orcid{0000-0002-7935-7445}\inst{3}
\and Ewine F. van Dishoeck\orcid{0000-0001-7591-1907}\inst{3,1} 
\and Danny Gasman\orcid{0000-0002-1257-7742}\inst{4}
\and Aditya M. Arabhavi\orcid{0000-0001-8407-4020}\inst{5}
\and Beno\^{i}t Tabone\orcid{0000-0002-1103-3225}\inst{6}
\and Thomas Henning\orcid{0000-0002-1493-300X}\inst{7} 
\and Inga Kamp\orcid{0000-0001-7455-5349}\inst{5} 
\and Alessio Caratti o Garatti\orcid{0000-0001-8876-6614}\inst{8,9}
\and Valentin Christiaens\orcid{0000-0002-0101-8814}\inst{4,10}
\and Pac\^{o}me Esteve\inst{6}
\and Manuel G\"udel\orcid{0000-0001-9818-0588}\inst{11,12}
\and Hyerin Jang\orcid{0000-0002-6592-690X}\inst{13}
\and Till Kaeufer\orcid{0000-0001-8240-978X}\inst{14}
\and Nicolas T. Kurtovic\orcid{0000-0002-2358-4796}\inst{1}  
\and Maria Morales-Calder\'on\orcid{0000-0001-9526-9499}\inst{15}
\and Giulia Perotti\orcid{0000-0002-8545-6175}\inst{16,7}
\and Kamber Schwarz\orcid{0000-0002-6429-9457}\inst{7}
\and Andrew D. Sellek\orcid{0000-0003-0330-1506}\inst{3}
\and Lucas M. Stapper\orcid{0000-0001-9524-3408}\inst{7}
\and Marissa Vlasblom\orcid{0000-0002-3135-2477}\inst{3}
\and L. B. F. M. Waters\orcid{0000-0002-5462-9387}\inst{13,17}
} 

\institute{
Max-Planck-Institut f\"ur Extraterrestrische Physik, Giessenbachstrasse 1, D-85748 Garching, Germany, 
\and Earth and Planets Laboratory, Carnegie Institution for Science, 5241 Broad Branch Road, NW, Washington, DC 20015, USA \email{sgrant@carnegiescience.edu} 
\and 
Leiden Observatory, Leiden University, P.O. Box 9513, 2300 RA Leiden, the Netherlands 
\and Institute of Astronomy, KU Leuven, Celestijnenlaan 200D, 3001 Leuven, Belgium
\and 
Kapteyn Astronomical Institute, Rijksuniversiteit Groningen, Postbus 800, 9700AV Groningen, The Netherlands
\and Universit\'e Paris-Saclay, CNRS, Institut d’Astrophysique Spatiale, 91405, Orsay, France
\and
Max-Planck-Institut f\"{u}r Astronomie (MPIA), K\"{o}nigstuhl 17, 69117 Heidelberg, Germany
\and INAF – Osservatorio Astronomico di Capodimonte, Salita Moiariello 16, 80131 Napoli, Italy
\and Dublin Institute for Advanced Studies, 31 Fitzwilliam Place, D02 XF86 Dublin, Ireland
\and STAR Institute, Universit\'e de Li\`ege, All\'ee du Six Ao\^ut 19c, 4000 Li\`ege, Belgium
\and Dept. of Astrophysics, University of Vienna, T\"urkenschanzstr. 17, A-1180 Vienna, Austria
\and ETH Z\"urich, Institute for Particle Physics and Astrophysics, Wolfgang-Pauli-Str. 27, 8093 Z\"urich, Switzerland
\and Department of Astrophysics/IMAPP, Radboud University, PO Box 9010, 6500 GL Nijmegen, The Netherlands
\and Department of Physics and Astronomy, University of Exeter, Exeter EX4 4QL, UK
\and Centro de Astrobiolog\'ia (CAB), CSIC-INTA, ESAC Campus, Camino Bajo del Castillo s/n, 28692 Villanueva de la Ca\~nada, Madrid, Spain
\and Niels Bohr Institute, University of Copenhagen, NBB BA2, Jagtvej 155A, 2200 Copenhagen, Denmark
\and SRON Netherlands Institute for Space Research, Niels Bohrweg 4, NL-2333 CA Leiden, the Netherlands
}

\abstract
{The chemical composition of the inner regions of disks around young stars will largely determine the properties of planets forming in these regions. Many disk physical processes drive the disk chemical evolution, some of which depend on and/or correlate with the stellar properties.}
{We aim to explore the connection between stellar properties and the inner disk chemistry in protoplanetary disks, as traced by mid-infrared spectroscopy.}
{We use JWST-MIRI observations of a large, diverse sample of sources to explore trends between the carbon-bearing molecule C$_2$H$_2$ and the oxygen-bearing molecule H$_2$O. Additionally, we calculate the average spectrum for the T Tauri (\mstar$>$0.2 \Msun) and very low-mass star (VLMS, \mstar$\leq$0.2 \Msun) samples from JWST-MIRI MRS data and use slab models to determine the properties of the average spectra in each subsample. }
{We find a significant anti-correlation between the flux ratio of C$_2$H$_2$/H$_2$O and the stellar luminosity. Disks around VLMS have significantly higher $F_{\rm{C_2H_2}}$/$F_{\rm{H_2O}}$ flux ratios than their higher-mass counterparts, driven by the generally weak H$_2$O and strong C$_2$H$_2$ in disks around low-mass hosts. We also explore trends with the strength of the 10 $\mu$m silicate feature, the stellar accretion rate, and the disk dust mass, all of which show correlations with $F_{\rm{C_2H_2}}$/$F_{\rm{H_2O}}$, which may be related to processes driving the carbon-enrichment in disks around VLMS, but also have degeneracies with system properties (i.e., the \mstar--\Mdot\ and \mstar--\mdisk\ relationships). Slab model fits to the average spectra show that H$_2$O emission in the VLMS sample is quite similar in temperature and column density to a warm ($\sim$600 K) H$_2$O component in the T Tauri spectrum, indicating that the high C/O gas phase ratio in these disks is not due to oxygen depletion alone. Instead, the presence of many hydrocarbons, including some with high column densities, suggests that carbon enhancement in the disks around VLMS is taking place.}
{The observed differences in the inner disk chemistry as a function of host properties are likely to be accounted for by differences in the disk temperatures, stellar radiation field, and the evolution of dust grains. }


\date{Received June 7th, 2025 / Accepted August 5th, 2025}

\maketitle

\section{Introduction}\label{sec: intro}
Understanding the main factors impacting the protoplanetary disk chemistry is of great importance for understanding the conditions during planet formation and therefore for constraining what material is available for nascent planets, both rocky and giant (e.g., \citealt{oberg_bergin21}). The evolution and structure of the dust disk, both vertically and radially, and the stellar luminosity, which controls the disk temperature, are expected to have a strong impact on the gaseous composition of the inner, terrestrial-planet-forming regions of disks. Infrared spectroscopy is a critical tool, not just for determining what molecules are present in these inner few au of protoplanetary disks, but also for determining the conditions of the gas and the processes that are at play in setting the chemistry. Observations with the Infrared Space Observatory, \textit{Spitzer}, and now JWST provide a unique window into these inner-disk regions, and the sensitivity and spectral resolution of JWST is ushering in new understandings of the first stages of planet formation. 

With large, diverse samples of protoplanetary disks observed in the mid-infrared, trends between the chemistry and the physical properties of these systems can be investigated, providing a way to determine what factors are driving the chemistry. Large samples observed with \textit{Spitzer} using the Infrared Spectrograph (IRS) provided great insight into the links between inner disk chemistry with overall disk properties (e.g., \citealt{pontoppidan14a}). For instance, a trend was found between the flux ratio of HCN/H$_2$O and the disk mass, potentially indicating that more massive disks lock up oxygen-rich ices in the cold outer regions, depriving the inner disk of H$_2$O enrichment \citep{najita13}. This conclusion was reinforced by trends found between the H$_2$O/HCN flux ratio and the disk dust radius, with smaller disks having more H$_2$O relative to HCN \citep{banzatti20}. Trends have also been found as a function of the stellar properties, largely the stellar mass and luminosity. \cite{pascucci09, pascucci13} found that cool stars (spectral type later than M5) have very different chemical signatures in the mid-infrared than earlier type stars (spectral type between K1 and M5), with the cool stars having higher C$_2$H$_2$ fluxes than HCN, where the opposite is true for earlier-type stars. These trends provided key insights into the impact of disk evolution and stellar properties on setting inner-disk chemistry and were only possible by access to large, diverse observational samples. 

JWST results are now building on the legacy of \textit{Spitzer} in the study of the inner regions of protoplanetary disks. The increased sensitivity and spectral resolution of JWST-MIRI, relative to \textit{Spitzer}-IRS, is allowing for the detection of isotopologues (e.g., \citealt{grant23a, tabone23, salyk25}), the de-blending of both ro-vibrational and pure rotational H$_2$O lines (e.g., \citealt{pontoppidan24a, gasman23b}), and detections of very weak emission to reveal previously unknown molecular content (e.g., \citealt{perotti23,arabhavi24a,arabhavi25a}). These advancements, along with analysis of increasingly large samples (e.g., \citealt{romero-mirza24b,arabhavi25b,arulanantham25,banzatti25a}), are transforming our understanding of inner disk chemistry by allowing for a better accounting of the chemical complexity, providing tighter constraints on the gas properties, and by allowing us to explore relationships with system properties.

In this work, we take a large sample of disks observed with JWST-MIRI MRS and find a strong anti-correlation between the C$_2$H$_2$/H$_2$O flux ratio and the stellar luminosity, spanning three orders of magnitude in stellar luminosity (and two orders of magnitude in stellar mass, including the transition from the stellar to sub-stellar regimes) and over three orders of magnitude in flux ratio. In Section~\ref{sec: obs and methods} we present the sample, observations, and methods for calculating line fluxes and dust properties. In Section~\ref{sec: results} we present the anti-correlations between $F_{\rm{C_2H_2}}$/$F_{\rm{H_2O}}$ and the stellar luminosity, accretion rate, disk dust mass, and the strength of the 10 \mic\ silicate feature and the average spectrum for the T Tauri and VLMS samples, as well as a slab model fit to those average spectra. We discuss these results and the degeneracies between many of the system parameters in Section~\ref{sec: discussion}. We provide a summary of our findings in Section~\ref{sec: summary}.

\section{Sample, observations, and methods}\label{sec: obs and methods}

\subsection{Sample}\label{subsec: sample}
Our sample comes from the Mid-INfrared Disk Survey (MINDS) JWST Guaranteed Time program (PID 1282, PI Henning, \citealt{henning24,kamp23}). The entire MINDS sample consists of 52 targets, spanning stellar masses from the sub-stellar brown dwarf regime to Herbig Ae stars and five debris disks. We do not include debris disks, highly inclined sources, Herbig Ae/Be systems, or sources dominated by PAH emission in our analysis. We also include the brown dwarf system TWA 27A/2MASS J12073346-3932539 from PID 1270, PI Birkmann; see \citealt{patapis25}). For this work,  ``T Tauri'' sample is taken to be objects with a stellar mass above 0.2 \Msun, while the ``VLMS'' is taken as those with host masses of 0.2 \Msun\ and below, including objects in the brown dwarf regime. Our sample thus consists of nine very low-mass stars and brown dwarfs (\mstar$\sim$0.02 to 0.16 \Msun, SpT from M4.5 to M9, see \citealt{arabhavi25b} for more details on this sample) and 25 T Tauri stars (\mstar$\sim$0.25 to 1.5 \Msun, SpT from M4 to G8). All of these sources were observed with the Mid-Infrared Instrument (MIRI), in the medium-resolution mode (MRS, \citealt{wright23}). This provides spectra from 4.9 to 28 \mic\ at a resolving power of $R\sim$1500-3500.

The stellar and disk properties for our sample are collected from the literature and provided in Table~\ref{tab: stellar props and line fluxes}. The accretion rate and disk dust masses are taken from the compilation of \cite{manara23}.

\begin{table*}
\centering
\caption{Stellar properties and line fluxes for our sample.} 
\label{tab: stellar props and line fluxes}
\begin{tabular}{|c|cccccc|cc|}
\hline \hline 
Target & $M_*$ & SpT & $L_*$ & $F_{9.8}$ & $log_{10}(\dot{M})$ & $M_{\rm{dust}}$ & $F_{\rm{C_2H_2}}$ & $F_{\rm{H_2O}}$\\
 &   [$M_{\odot}$] & & [$L_{\odot}$] &  &[$M_{\odot}/yr$] & [$M_{\oplus}$]  & [$\times$10$^{-14}\ \rm{erg\ s^{-1}\ cm^{-2}}$] & [$\times$10$^{-14}\ \rm{erg\ s^{-1}\ cm^{-2}}$] \\
\hline
J04390163+2336029	&	0.08		&	M6	$^{a}$	&	0.1		&	1.43	&	-9.71		&	3.0		&	0.38 $\pm$ 0.01	&	0.04 $\pm$ 0.01  \\
J11071668-7735532	&	0.05		&	M7.74	$^{a}$	&	0.02	$^{b}$	&	1.69	&	-11.69	$^{c}$	&			&	1.19 $\pm$ 0.04	&	$<$ 0.11  \\
J11071860-7732516	&	0.08		&	M5.5		&	0.04		&	1.0	&	-10.4		&	0.38		&	0.49 $\pm$ 0.01	&	0.06 $\pm$ 0.01  \\
J11074245-7733593	&	0.12		&	M5.5		&	0.06		&	1.18	&	-9.62		&	0.89		&	1.32 $\pm$ 0.02	&	$<$ 0.06 $^{*}$ \\
J11082650-7715550	&	0.07		&	M5.5		&	0.02		&	1.0	&	-10.67		&	0.19		&	0.71 $\pm$ 0.02	&	$<$ 0.05 $^{*}$ \\
J11085090-7625135	&	0.08		&	M5.5		&	0.04		&	1.13	&	-10.27		&	0.18		&	0.5 $\pm$ 0.02	&	$<$ 0.05 $^{*}$ \\
J12073346-3932539	&	0.02	$^{d}$	&	M9	$^{a}$	&	0.01	$^{d}$	&	1.0	&	-11.23	$^{d}$	&	0.1	$^{a}$	&	0.14 $\pm$ 0.0	&	$<$ 0.01  \\
J15582981-2310077	&	0.16		&	M4.5		&	0.04		&	1.33	&	-9.05		&	1.19		&	2.17 $\pm$ 0.01	&	0.08 $\pm$ 0.01  \\
J16053215-1933159	&	0.16		&	M4.5	$^{a}$	&	0.03	$^{e}$	&	1.0	&	-9.4	$^{f}$	&	0.14		&	4.5 $\pm$ 0.02	&	$<$ 0.05 $^{*}$ \\
AA Tau	&	0.68		&	M0.6		&	0.75		&	1.48	&	-7.64		&	36.0		&	2.21 $\pm$ 0.15	&	5.59 $\pm$ 0.17  \\
BP Tau	&	0.49		&	M0.5		&	0.98		&	2.1	&	-7.29		&	25.01		&	$<$ 0.33	&	3.32 $\pm$ 0.12  \\
CX Tau	&	0.33		&	M2.5		&	0.34	$^{g}$	&	1.52	&	-9.59	$^{h}$	&	4.21		&	0.39 $\pm$ 0.03	&	0.24 $\pm$ 0.03  \\
CY Tau	&	0.42		&	M2.3		&	0.36		&	1.08	&	-8.2		&	35.81		&	1.67 $\pm$ 0.04	&	0.55 $\pm$ 0.04  \\
DF Tau	&	0.39		&	M2.7		&	0.59	$^{i}$	&	1.21	&	-7.77	$^{i}$	&	1.8		&	22.24 $\pm$ 1.42	&	13.33 $\pm$ 1.47  \\
DL Tau	&	1.07		&	K5.5		&	1.49		&	1.13	&	-7.19		&	123.4		&	14.28 $\pm$ 0.13	&	2.58 $\pm$ 0.14  \\
DM Tau	&	0.29		&	M3		&	0.24		&	1.41	&	-7.99		&	52.42		&	$<$ 0.06 $^{*}$	&	$<$ 0.06 $^{*}$ \\
DN Tau	&	0.53		&	M0.3		&	0.7		&	1.26	&	-8.18		&	41.41		&	$<$ 0.23 $^{*}$	&	0.36 $\pm$ 0.08  \\
DR Tau	&	0.83		&	K6		&	1.9		&	1.38	&	-6.71		&	70.44		&	14.21 $\pm$ 1.11	&	29.15 $\pm$ 1.24  \\
FT Tau	&	0.3		&	M2.8		&	0.45		&	1.63	&	-8.92	$^{j}$	&	42.99		&	1.24 $\pm$ 0.12	&	2.11 $\pm$ 0.12  \\
GW Lup	&	0.46		&	M1.5		&	0.33		&	1.6	&	-9.03		&	50.06		&	1.36 $\pm$ 0.07	&	0.72 $\pm$ 0.07  \\
IM Lup	&	1.09		&	K5		&	2.51		&	1.29	&	-7.85		&	137.22		&	3.04 $\pm$ 0.14	&	3.01 $\pm$ 0.15  \\
LkCa15	&	0.7		&	K5.5		&	1.1		&	3.02	&	-7.94		&	88.67		&	$<$ 0.09 $^{*}$	&	0.25 $\pm$ 0.03  \\
PDS 70	&	0.76		&	K7	$^{k}$	&	0.38		&	3.82	&	-10.26	$^{l}$	&	21.83	$^{m}$	&	$<$ 0.04 $^{*}$	&	0.12 $\pm$ 0.01  \\
RNO 90	&	1.5		&	G8	$^{n}$	&	2.69		&	1.47	&	-7.26	$^{n}$	&			&	10.19 $\pm$ 0.39	&	18.64 $\pm$ 0.46  \\
RW Aur	&	1.5		&	K0		&	2.51		&	1.25	&	-7.07		&	19.72		&	15.41 $\pm$ 0.69	&	24.19 $\pm$ 0.82  \\
SY Cha	&	1.12		&	K7		&	0.55		&	1.82	&	-9.18		&	1.33		&	0.66 $\pm$ 0.1	&	1.38 $\pm$ 0.1  \\
Sz 50	&	0.25		&	M4		&	0.69		&	1.26	&	-8.79		&	11.96		&	0.57 $\pm$ 0.08	&	0.59 $\pm$ 0.09  \\
Sz 98	&	0.55		&	K7		&	1.53		&	1.91	&	-7.44		&	75.85		&	$<$ 0.63 $^{*}$	&	4.33 $\pm$ 0.22  \\
TW Hya	&	0.61		&	M0.5	$^{o}$	&	0.23	$^{o}$	&	1.96	&	-8.52	$^{p}$	&	70.12	$^{q}$	&	$<$ 0.11	&	0.27 $\pm$ 0.04  \\
V1094 Sco	&	0.64		&	K6		&	1.21		&	1.14	&	-7.88		&	135.05		&	1.79 $\pm$ 0.01	&	0.32 $\pm$ 0.01  \\
VW Cha	&	0.7		&	K7		&	2.31		&	1.53	&	-7.38		&	16.59		&	$<$ 1.73 $^{*}$	&	32.57 $\pm$ 0.71  \\
WA Oph 6	&	0.63		&	K7	$^{o}$	&	0.76	$^{o}$	&	1.22	&	-7.34	$^{o}$	&	46.6	$^{r}$	&	2.47 $\pm$ 0.16	&	1.81 $\pm$ 0.16  \\
WX Cha	&	0.49		&	M0.5		&	0.86		&	1.78	&	-6.73		&	7.78		&	4.2 $\pm$ 0.32	&	6.91 $\pm$ 0.35  \\
XX Cha	&	0.25		&	M3.5		&	0.41		&	1.4	&	-7.17		&	8.03		&	4.56 $\pm$ 0.17	&	2.44 $\pm$ 0.18  \\
\hline
\end{tabular}
\tablefoot{Upper limits are given as 3$\sigma$. Fluxes marked with an $^{*}$ indicates that the non-detection came from by-eye inspection of line-rich or noisy spectra. The target names for the first nine targets are the 2MASS names. Values for \mstar, SpT, \Lstar, $log_{10}(\dot{M})$, and $M_{\rm{dust}}$ are compiled from \cite{manara23} unless otherwise noted. $^{a}$ \cite{arabhavi25b},  $^{b}$ \cite{luhman07d}, $^{c}$ \cite{manara16}, $^{d}$ \cite{manjavacas24},  $^{e}$ \cite{testi22}, $^{f}$ \cite{franceschi24}, $^{g}$ \cite{herczeg&hillenbrand14}, $^{h}$ \cite{vlasblom25a}, $^{i}$ \cite{grant24b}, $^{j}$ \cite{gangi22}, $^{k}$ \cite{pecaut_mamajek16}, $^{l}$ \cite{haffert19}, $^{m}$ calculated using the 1.33 mm flux from \cite{facchini21} assuming $T_{dust}$=20 K and $\kappa_{\nu}$=2.254 cm$^{-2}$ g$^{-1}$, $^{n}$ \cite{fang23}, $^{o}$ \cite{pascucci20}, $^{p}$ \cite{wendeborn24a}, $^{q}$ \cite{das24}, $^{r}$ \cite{brown-sevilla21}}
\end{table*}

\subsection{JWST observations and data reduction}\label{subsec: observations}

The entire sample has been reduced using the standard pipeline reduction (version 1.16.1; \citealt{bushouse_1.16.1}) and pmap 1315. Aperture photometry with an aperture size of 2$\times$ the full width at half maximum was used to extract the spectra. Residual
fringes were removed using the default pipeline. 

Continuum-subtraction is done following the methods of \cite{temmink24a}. Briefly, this is done via an iterative fitting with the continuum being fit using a Savitzky-Golay filter with a third-order polynomial. Emission lines above 2$\sigma$ above the continuum are masked so as to not skew the continuum estimation. The continuum is then subtracted, and all downward spikes more than 3$\sigma$ below
the continuum are masked. Finally, the baseline is determined using PyBaselines \citep{erb22}. Molecular pseudo-continuum is present in all of the VLMS. For these sources, the wavelength ranges where the pseudo-continua are obviously present are masked by eye from the continuum determination.

\begin{figure*}[h]
    \centering
    \includegraphics[scale=0.55]{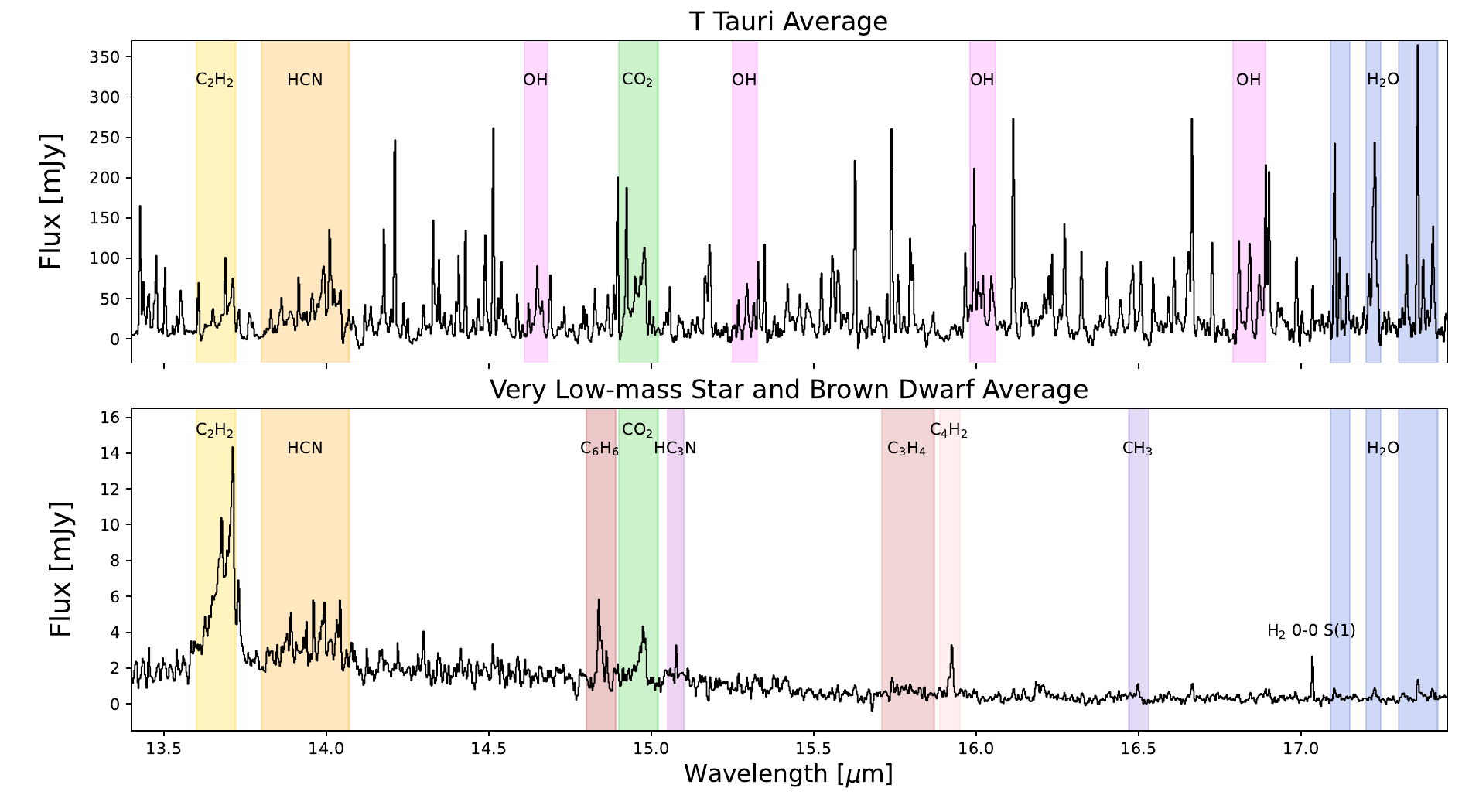}
    \caption{The average T Tauri (top) and VLMS (bottom) sample JWST spectra. The detected molecular species are highlighted. The C$_2$H$_2$ and H$_2$O regions (yellow and blue, respectively) show the regions over which the fluxes are integrated. Most of the unlabeled lines in the T Tauri average are various H$_2$O transitions.   }
    \label{fig: average spectra}
\end{figure*}

\subsection{Line fluxes}\label{subsec: line fluxes}

The H$_2$O flux is determined by integrating the spectrum over three windows centered on lines with upper energy levels of 2400 to 6000 K and Einstein A coefficients of 1 to 42 s$^{-1}$ in the 17 \mic\ range: 17.09--17.15 \mic, 17.2--17.245 \mic, and 17.3--17.42 \mic. These transitions, and thus the integrated flux that we measure,  largely trace a ``warm'' H$_2$O component, with a temperature of $\sim$400 K. However, H$_2$O has been found to have gradients in temperature in disks (e.g., \citealt{temmink24b,gasman23b,banzatti23b,grant24b,munoz-romero24a}), therefore we are only tracing a portion of the H$_2$O in these systems and some may have different reservoirs of hot and cold water. However, these lines are more commonly detected for the VLMS sample than hot ro-vibrational lines at shorter wavelengths and are less blended with hydrocarbon features between $\sim$12 and 16 \mic. Finally, the noise level is too high to access the very cold components at $\sim$24 \mic\ \citep{arabhavi25a}.

In H$_2$O-rich spectra, H$_2$O lines can contaminate the C$_2$H$_2$ $Q-$branch at 13.7 \mic. Therefore, we follow the methods of \cite{banzatti20} to remove the H$_2$O contribution before determining the C$_2$H$_2$ flux. We take a local thermodynamic equilibrium H$_2$O slab model with a temperature of 600 K and a column density of 10$^{18}$ cm$^{-2}$ (properties that have been found for H$_2$O lines near the C$_2$H$_2$ feature, e.g., \citealt{grant23a,gasman23b}), scale it to match the continuum and peak fluxes of water lines in two windows (13.415 to 13.445 \mic\ and 14.19 to 14.35 \mic) close in wavelength to the C$_2$H$_2$ $Q$-branch peak. This H$_2$O model is then subtracted from the observed spectrum before the integrated C$_2$H$_2$ flux is measured. We note that removing the H$_2$O emission is not necessary for the VLMS, as the H$_2$O emission is very weak, if present at all, and any fit for H$_2$O will be greatly contaminated by the strong C$_2$H$_2$ and HCN emission. The C$_2$H$_2$ flux is integrated from 13.60 to 13.72 \mic.

To determine whether H$_2$O and C$_2$H$_2$ are detected requires an understanding of the noise level in the spectrum. This can be very challenging in the line-rich MIRI spectra, as there is little pure continuum on which to measure the noise level. Based on model spectra of the common molecular species found in the 13 to 18 \mic\ wavelength range, we select wavelength regions which contain the least molecular emission in which to determine the noise (from 15.895 to 15.91 \mic\ for the T Tauris and 16.375 and 16.395 \mic\ for the VLMS). The species is considered detected if the line flux is greater than 3$\sigma$. However, there are several cases in which the spectra are very line-rich or are noisy, leading to 3$\sigma$ ``detections'' that are not reliable when inspected by eye; for instance, the integrated flux over the C$_2$H$_2$ feature is high but the region has residual water lines and/or no discernible C$_2$H$_2$ $Q-$branch. In these cases we consider these non-detections, but provide the 3$\sigma$ upper limits in Table~\ref{tab: stellar props and line fluxes}. Finally, while we use the 3$\sigma$ threshold, confirmed by visual inspection, to determine whether the emission is detected in this work, H$_2$O has been detected in most of the VLMS through cross-correlation, either at shorter wavelengths or in the rotational lines that we analyze here. We refer the reader to \cite{arabhavi25a} for details on these H$_2$O detections in the VLMS sample.

\begin{figure*}[h]
    \centering
    \includegraphics[scale=0.8]{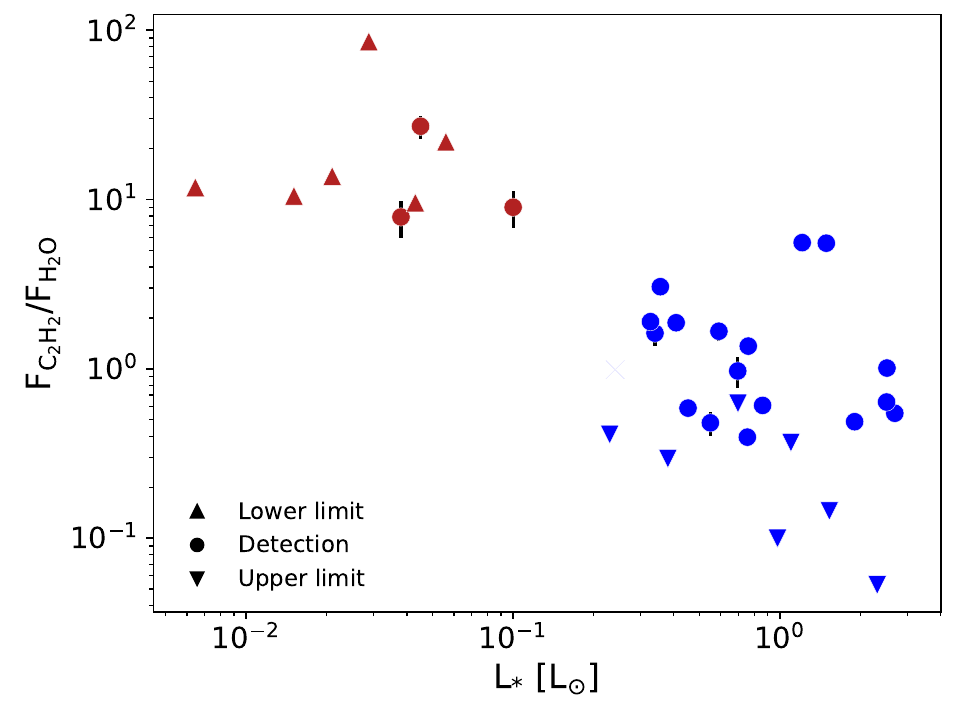}
    \caption{The relationship between the flux ratio of C$_2$H$_2$ to H$_2$O as a function of stellar luminosity. Objects with \mstar$>$0.2 \Msun\ are colored in blue and represent our T Tauri sample, while those with stellar masses below 0.2 \Msun\ are our VLMS sample and are shown in red. The two outliers around \Lstar$\sim$1 \Lsun\ are DL Tau and V1094 Sco, which will be analyzed in detail in Tabone et al. (in prep.). This trend is statistically significant with a $p-$value of 1.3$\times$10$^{-7}$ and a correlation coefficient of -0.77. Upper/lower limits (downward/upward triangles) are the 3$\sigma$ limits. Error bars are smaller than the points for most targets. }
    \label{fig: C2H2 H2O Lstar}
\end{figure*}

\subsection{10 $\mu$m silicate feature strength}
In order to investigate the connection between the gas properties and the (infrared) dust properties, we calculate the strength of the 10 \mic\ silicate feature following the methods of e.g., \cite{vanboekel03, kessler-silacci06}: $F_{9.8}$ = 1 + ($f_{9.8, cs}$/ $<f_c>$), where $f_{9.8, cs}$ is the spectrum after subtracting the continuum that is determined by two anchor points, one from 6.8 to 7.5 \mic\ and the other from 12.5 to 13.5 \mic, and $<f_c>$ is the mean of the continuum. For four of the VLMS (2MASS J16053215-1933159, 2MASS J11071860-7732516, 2MASS J11082650-7715550, and 2MASS J12073346-3932539), there is either no 10 \mic\ emission or any emission around 10 \mic\ is coming at least partially from optically thick C$_2$H$_4$ (\citealt{arabhavi24a, arabhavi25b}), therefore for these sources we adopt a band strength of 1, representative of no silicate emission.

\section{Results}\label{sec: results}

\subsection{Average spectrum}\label{subsec: average spec}

To explore the properties of the T Tauri and VLMS samples, we take the average spectrum of each sample to perform a spectral fit. Each spectrum is scaled to a common distance of 120 pc before the spectra are averaged in each wavelength bin. The average spectra of the T Tauri and VLMS samples are presented in Figure~\ref{fig: average spectra}. This figure starkly highlights the difference in the molecular emission between the two samples. H$_2$O lines are the strongest features in the averaged T Tauri spectrum, followed by the CO$_2$ and HCN $Q$-branches, OH, and finally C$_2$H$_2$. By comparison, the averaged VLMS spectrum is dominated by the bright C$_2$H$_2$ $Q$-branch, followed by C$_6$H$_6$, HCN, CO$_2$, C$_4$H$_2$, HC$_3$N, $^{13}$CO$_2$, and finally H$_2$O being the weakest feature.

\subsection{The flux ratio between C$_2$H$_2$ and H$_2$O}\label{subsec: C2H2 H2O}

\begin{figure*}[h]
    \centering
    \includegraphics[scale=0.5]{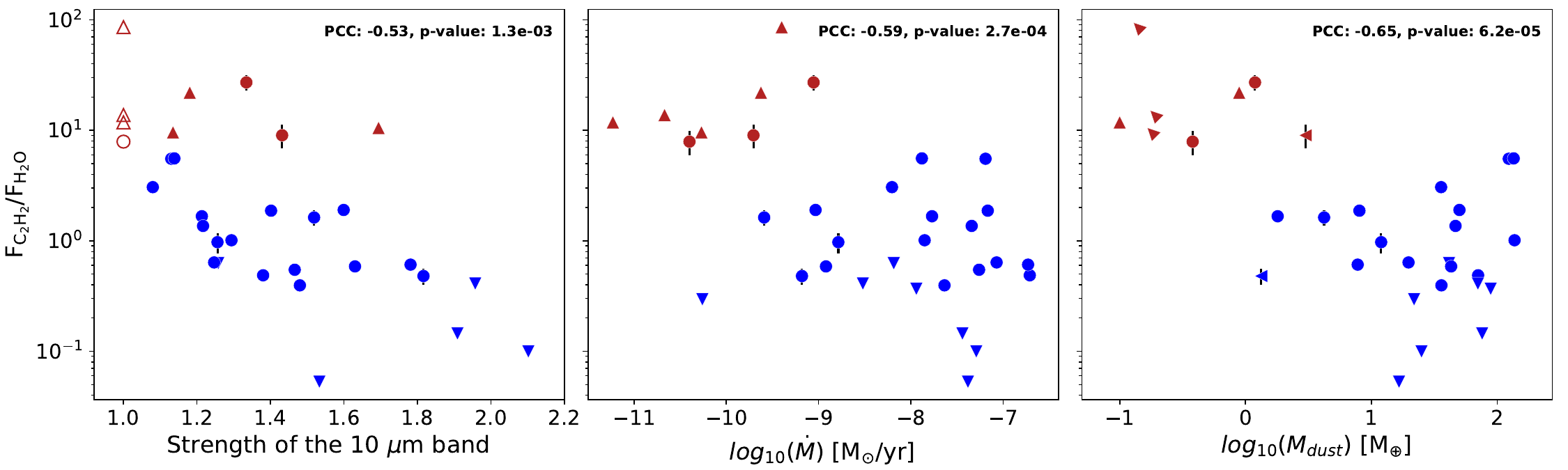}
    \caption{Left: $F_{\rm{C_2H_2}}$/$F_{\rm{H_2O}}$ as a function of the strength of the 10 \mic\ silicate feature (stronger silicate features have higher values). For four of the VLMS, there is either no 10 \mic\ emission or the emission is coming at least partially from C$_2$H$_4$, therefore we adopt feature strength of 1 for these sources (denoted by open points). Two outliers at a 10 \mic\ band strength of 3 and 3.8 are the transitional disks LkCa 15 and PDS 70 and have been removed for clarity. Middle: The relationship between $F_{\rm{C_2H_2}}$/$F_{\rm{H_2O}}$ and \Mdot. Right: The relationship between $F_{\rm{C_2H_2}}$/$F_{\rm{H_2O}}$ and \Mdust. The PCCs and $p-$values can be found for each panel. All of the relationships are statistically significant ($p-$value$<$0.05); however the correlations are not as strong as the $F_{\rm{C_2H_2}}$/$F_{\rm{H_2O}}$ vs. stellar luminosity relationship. Rotated triangular markers for the VLMS sample indicate lower limits on $F_{\rm{C_2H_2}}$/$F_{\rm{H_2O}}$ and upper limits on \Mdust. Error bars are smaller than the points for most targets. }
    \label{fig: C2H2 H2O Mstar Mdot Mdust}
\end{figure*}

\begin{figure*}[h]
    \centering
    \includegraphics[scale=0.48]{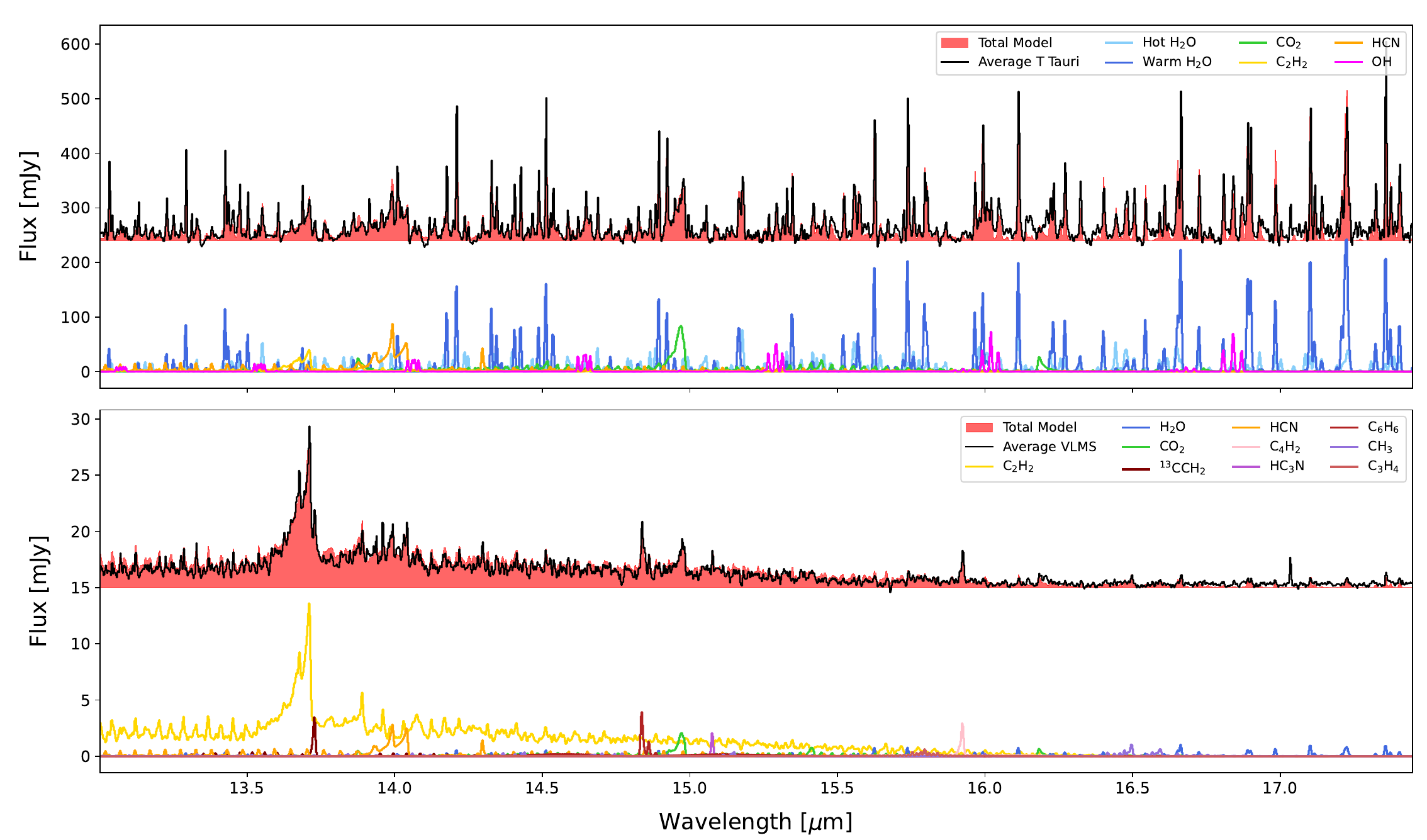}
    \caption{The average spectra for disks in our T Tauri sample (top) and VLMS sample (bottom) in black, compared to the best-fit model in red. The model components are shown below each for reference.}
    \label{fig: avg fit}
\end{figure*}

\begin{figure*}[h]
    \centering
    \includegraphics[scale=0.55]{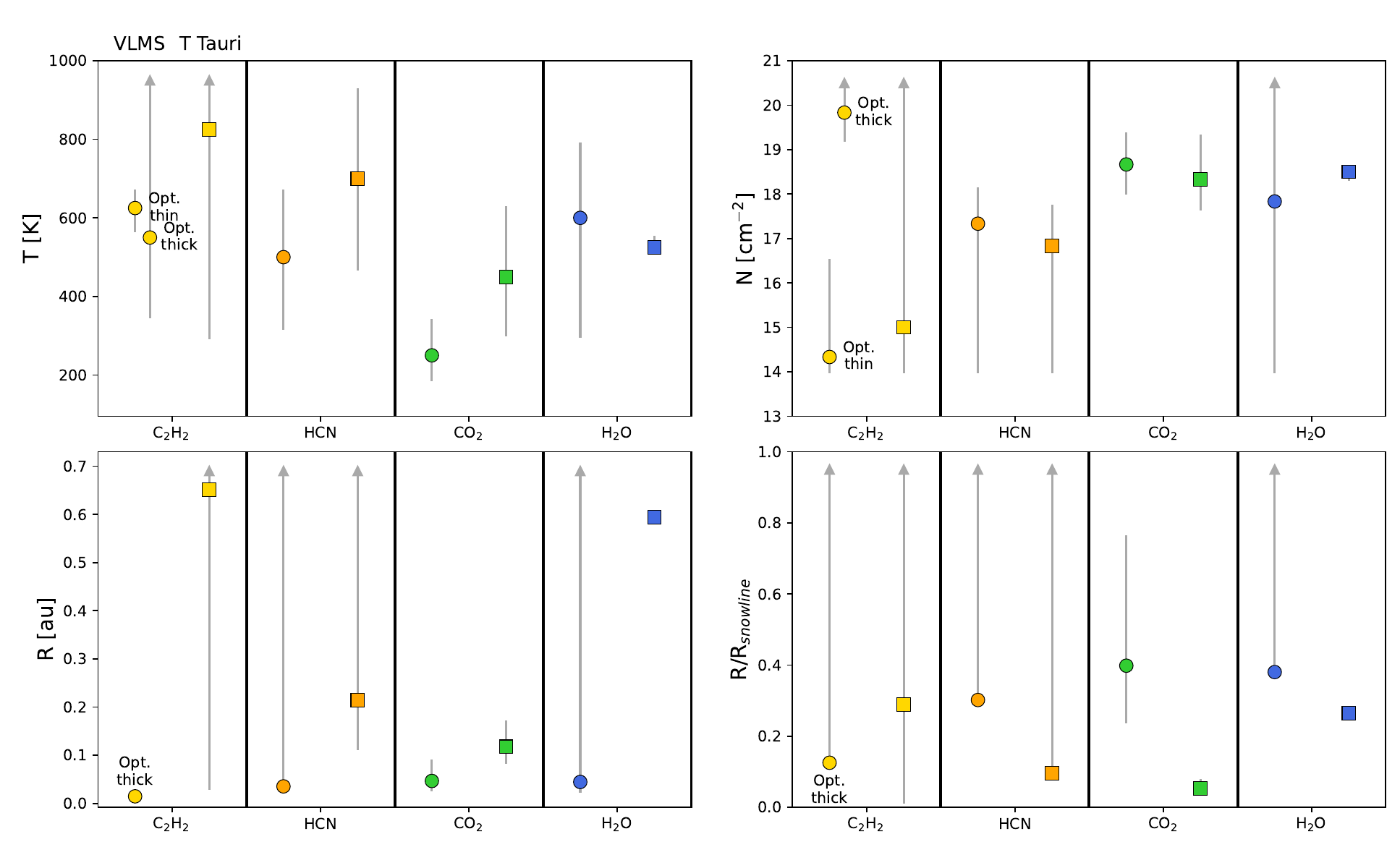}
    \caption{The best-fit slab model parameters for the average VLMS spectrum (left circles) and T Tauri spectrum (right squares) for the different molecules. The H$_2$O component for the T Tauri spectrum is the intermediate ($T\sim600$) component. Temperature and column density are shown on the top on the left and right, respectively. The equivalent emitting radius is shown on the bottom left and that radius normalized to the H$_2$O snowline for each subsample is shown on the bottom right. Only the optically thick C$_2$H$_2$ component is shown for the VLMS in the radii plots, as the radius is unconstrained in the optically thin case. Error bars are determined from the 1$\sigma$ confidence intervals and in some cases are degenerate between parameters. For example, in the case of optically thin emission, the column density and equivalent emitting radii are degenerate. See Figures~\ref{fig: chi2 ttauri} and \ref{fig: chi2 vlms} for the $\chi^2$ maps for the T Tauri and VLMS fits, respectively.}
    \label{fig: avg params}
\end{figure*}

The anti-correlation between $F_{\rm{C_2H_2}}$/$F_{\rm{H_2O}}$ (identically $L_{\rm{C_2H_2}}$/$L_{\rm{H_2O}}$) and the stellar luminosity can be seen in Figure~\ref{fig: C2H2 H2O Lstar}. This correlation spans over three orders of magnitude in stellar luminosity (almost two orders of magnitude in stellar mass) and over three orders of magnitude in $F_{\rm{C_2H_2}}$/$F_{\rm{H_2O}}$. We find that the absolute flux of H$_2$O decreases with decreasing stellar luminosity (Pearson correlation coefficient, PCC=0.84, $p-$value=6$\times$10$^{-10}$, but the C$_2$H$_2$ flux does not; instead, the C$_2$H$_2$ line-to-continuum ratio increases with decreasing stellar luminosity (PCC=-0.65, $p-$value=2.7$\times$10$^{-5}$; see Figure~\ref{fig: absolute fluxes} in Appendix~\ref{sec: absolute fluxes} and also Figure 8 in \citealt{arabhavi25a}). Therefore, the strong correlation that we find between $F_{\rm{C_2H_2}}$/$F_{\rm{H_2O}}$ and \Lstar\ (PCC=-0.77, $p-$value=1.3$\times$10$^{-7}$)
is due to a combination of weak H$_2$O and strong C$_2$H$_2$ features in the very low-mass objects and vice versa for the T Tauri stars. While there is the overall decrease in $F_{\rm{C_2H_2}}$/$F_{\rm{H_2O}}$ with increasing stellar luminosity, even at a given stellar luminosity in the T Tauri sample, the spread in line flux ratio is around $\sim$50. We note that we do not include the accretion luminosity in the luminosity plotted in Figure~\ref{fig: C2H2 H2O Lstar} because the stellar luminosity is much higher (\Lacc$\lesssim$0.1\Lstar\ or even \Lacc$\lesssim$0.01\Lstar\ for the range of stellar luminosities in our sample; e.g., \citealt{mendigutia15a,alcala17,manara17}); however, we note that this would only strengthen the correlation, if only modestly.

Additional correlations are seen with other system properties, including the stellar mass, accretion rate and disk dust mass (the latter two are presented in Figure~\ref{fig: C2H2 H2O Mstar Mdot Mdust}). Given the correlations between stellar luminosity and mass and the correlations of the other properties with stellar mass (\mstar--\Mdot\ as in e.g., \citealt{hillenbrand92,hartmann16,grant22,manara23} and \mstar--\Mdust\ as in e.g., \citealt{andrews13,pascucci16,manara23}), this is not surprising. The trends seen with these system parameters are not as strong as the trend with stellar luminosity. It would also be interesting to put the JWST results into context with the outer disk information; however, high resolution (sub-)millimeter observations of the VLMS sample are currently lacking, therefore we cannot explore any potential correlations with $R_{\rm{disk}}$ (however, see \citealt{arulanantham25} for the JWST Disk Infrared Spectral
Chemistry Survey sample of T Tauri objects). We discuss the potential role of radial drift further in Section~\ref{sec: discussion}. 

In Figure~\ref{fig: C2H2 H2O Mstar Mdot Mdust}, we also show the relationship between $F_{\rm{C_2H_2}}$/$F_{\rm{H_2O}}$ and the strength of the 10 \mic\ silicate feature. An overall negative correlation is observed, i.e. targets with weaker silicate features are more likely to have stronger C$_2$H$_2$ emission relative to H$_2$O. Interestingly, none of the sources with a 10 \mic\ band strength of greater than 1.85 have C$_2$H$_2$ detections. The potential underlying connection is discussed in Section~\ref{subsec: dust evolution}. Correlations between the parameters discussed above are also provided relative to the absolute fluxes and to the line-to-continuum ratios in Appendix \ref{sec: absolute fluxes}, where the strongest correlation is between the H$_2$O flux and the accretion rate, as previously found (e.g., \citealt{banzatti20}).

\subsection{Fitting the average spectra}\label{subsec: fitting}

In order to interpret the average properties, we use 0D LTE slab models (see \citealt{grant23a, tabone23} for details) to reproduce the average spectra using only three free parameters: the column density ($N$), temperature ($T$), and the emitting area, characterized by a disk with an emitting radius ($R$). In order to fit each spectrum, we follow the method of \cite{grant23a}. Briefly, the slab model spectra are calculated with a Gaussian line profile with a full width at half maximum of
$\Delta$V = 4.7 km s$^{-1}$ ($\sigma$ = 2 km s$^{-1}$) at a spectral resolving power of 2300 to match the resolution of MIRI MRS in the $\sim$13 to 17 \mic\ wavelength region. Then the model is sampled on the same wavelength grid as the data using \texttt{SpectRes} \citep{spectres}. A grid of models is calculated for each molecular species, with $N$ from 10$^{14}$ to 10$^{22}$ cm$^{-2}$ in steps of 0.166 in log$_{10}$-space, $T$ from 100 to 1500 K in steps of 25 K, and the emitting area is varied by changing the radius from 0.001 to 10 au in steps of 0.02 in log$_{10}$-space. One molecule is fit at a time, selecting wavelength regions that are relatively devoid of emission from other species. Then, the best-fit model is subtracted off and the next species is fit. This iterative approach has been shown to produce consistent results with analysis performed doing a simultaneous fitting of all molecular species (e.g., \citealt{grant23a} and \citealt{kaeufer24a}). Two H$_2$O components, one hot component with $T$$>$1000 K and one intermediate component with $T$$\sim$600 K, are needed to reproduce the lines in the 13-17.5 \mic\ wavelength range for the average T Tauri spectrum, in line with works on individual sources (e.g., \citealt{grant24b, temmink24b,romero-mirza24b,banzatti25a}). Similarly, two components are needed to fit the C$_2$H$_2$ in the VLMS spectrum in order to reproduce the molecular pseudo-continuum \citep{tabone23,arabhavi24a,kanwar24b}. The best-fit model is shown in Figure~\ref{fig: avg fit} and the best-fit properties for the molecules that are present in both the average VLMS and T Tauri spectra are presented in Figure~\ref{fig: avg params}. The $\chi^2$ maps are presented in Figure~\ref{fig: chi2 ttauri} and \ref{fig: chi2 vlms} for the T Tauri and VLMS spectra, respectively. 

While some of the molecules are not well constrained, there are some interesting conclusions we can draw from the average-fit properties. Notably, the column densities for HCN, CO$_2$, and H$_2$O (the intermediate component for the T Tauri spectrum) are consistent between the VLMS and the T Tauris. While the equivalent emitting radii are largely unconstrained due to the degeneracy between radii and column density in the optically thin regime, the radii do tend to smaller values in the VLMS compared to the T Tauris.
Given different snowline locations in disks around VLMS relative to T Tauri stars, we determine water snowline location for each subsample in order to normalize the emitting radii in the bottom right panel of Figure~\ref{fig: avg params}. We use Equation 2 in \cite{mulders15} to determine the snowline location, which is based on the 3D radiative transfer models of \cite{min11}, using the average stellar mass and accretion rate for each subsample (\mstar$\sim$0.6 \Msun\ and \Mdot$\sim$10$^{-7.75}$ \msunyr\ for the T Tauri sample and \mstar$\sim$0.08 \Msun\ and \Mdot$\sim$10$^{-10}$ \msunyr\ for the VLMS sample). This results in an H$_2$O snowline location of 2.25 and 0.12 au for the T Tauris and VLMS, respectively. With this normalization, the gas properties of the VLMS tend to have larger emitting radii than their higher-mass counterparts. One of the most well-constrained molecules in each case is CO$_2$, which is optically thick for both samples and has a cooler temperature than the other molecules. Our CO$_2$ and C$_2$H$_2$ temperatures are colder than what was found in the JDISCS sample of \cite{arulanantham25}. Although it will be necessary to model each spectrum in order to get a more global picture of the diversity in the molecular properties (as was done in \citealt{arulanantham25}), the average fit already provides interesting initial clues into the similarities and differences between the two samples. 

While not shown in Figure~\ref{fig: avg params}, the hydrocarbons besides C$_2$H$_2$ in the VLMS spectra tend to be cold ($<$300 K) and in the optically thin regime with column densities $\lesssim$10$^{-17.5}$ cm$^{-2}$ (see Figure~\ref{fig: chi2 vlms}). The difference in temperature between C$_2$H$_2$ and the other hydrocarbons is likely due at least in part to the high abundances of C$_2$H$_2$. The optically thick component likely originates in the very inner disk, given its small emitting area, seen now in several sources (\citealt{tabone23,arabhavi24a,kanwar24b,long25}). Based on the 1$\sigma$ contours for the equivalent emitting radius, the maximum radius for the optically thick C$_2$H$_2$ component is 2.5 times smaller than the minimum radius for the optically thin component (0.027 vs 0.067 au, respectively). Although we cannot put good constraints on the emitting area of the optically thin hydrocarbon emission (see Figure~\ref{fig: chi2 vlms}), the fact that the optically thin C$_2$H$_2$ component is hotter than the other more complex hydrocarbons could come from it emitting at higher layers in the disk atmosphere (see Appendix D in \citealt{kanwar24b}) or slightly closer to the star at higher temperatures. The latter could point to a radial change in the gas phase C/O ratio, wherein C$_2$H$_2$ can form at lower C/O ratios but the more complex hydrocarbons need higher values to form. Additional 2D thermochemical modeling, including non-LTE effects, will be useful to investigate the temperatures determined from the observations.

\section{Discussion}\label{sec: discussion}

We find a strong anti-correlation between the flux ratio of C$_2$H$_2$ to H$_2$O and the stellar luminosity. Most of the targets fall within the $F_{\rm{C_2H_2}}$/$F_{\rm{H_2O}}$ ratio found from the T Tauri modeling work of \cite{anderson21} (see their Figure 11, although we note that we do not use the exact same wavelength range to determine the fluxes). T Tauri objects with an observed $F_{\rm{C_2H_2}}$/$F_{\rm{H_2O}}$$\lesssim$2 would be most consistent with an oxygen-rich volatile content and a C/O ratio below 0.57, based on the \cite{anderson21} models. On the other hand, objects with host masses below $\sim$0.2 \Msun\ have significantly higher $F_{\rm{C_2H_2}}$/$F_{\rm{H_2O}}$ ratios than their higher-mass counterparts, stemming from their decreased H$_2$O fluxes and increased C$_2$H$_2$ fluxes. This work builds upon trends seen with \textit{Spitzer} observations by \cite{pascucci13} and initial results with JWST (see e.g., \citealt{kamp23,long25,arabhavi25b}). 

Many factors can influence both the composition of disks and how much of the composition can actually be observed. Additionally, many of these factors can be related, making it non-trivial to determine what is driving the observed correlations in these samples. With this in mind, we explore the correlations that we have found and discuss what physical processes may be dominating the large difference in chemical signatures between VLMS and T Tauris.

\subsection{Carbon enrichment or oxygen depletion in the VLMS disks?}

The hydrocarbon-dominated average VLMS spectrum in Figure~\ref{fig: avg fit} highlights the rich carbon chemistry that has been found in other VLMS with JWST thus far (e.g., \citealt{tabone23,arabhavi24a,kanwar24b, long25,morales-calderon25_submitted}). \cite{long25} use the models of \cite{najita11}, paired with the column density ratio of C$_2$H$_2$ to CO$_2$, to determine the C/O ratio in their carbon-rich disk. Doing the same for the average fit spectra, the average VLMS has a C/O of $\sim$1. It is worth investigating whether the high C/O ratio inferred for these disks is due to carbon enhancement and/or oxygen depletion. 

The high C$_2$H$_2$ column densities derived (e.g., \citealt{tabone23} and upper right panel of Figure~\ref{fig: avg params}) and the C$_2$H$_2$/HCN findings of \cite{pascucci09} point to carbon enhancement in the VLMS disks. Additionally, the presence of large hydrocarbon chains, including C$_2$H$_6$, C$_3$H$_4$, C$_4$H$_2$, and C$_6$H$_6$, is pointing to very abundant gas-phase carbon and a gas-phase C/O$>$1. We note that C$_2$H$_4$ and C$_2$H$_6$ are present in a few sources, but at shorter wavelengths than we analyze here; see \cite{arabhavi25b} for more details.

An interesting finding pointing away from at least some oxygen depletion being the cause of the high C/O is the common presence of CO$_2$ and $^{13}$CO$_2$ in these disks (e.g., \citealt{tabone23, arabhavi24a, arabhavi25b}). Additionally, it has now been found that H$_2$O is in fact present in these disks, albeit at very low flux levels \citep{arabhavi25a}, in line with the decrease in stellar luminosity (Figure~\ref{fig: absolute fluxes}) and therefore emitting area, especially as the H$_2$O snowline is very close to the central object in these disks. This will particularly impact the hottest H$_2$O, mostly in the ro-vibrational lines at shorter wavelengths as the emitting area is very radially compact (see Figure 9 in \citealt{arabhavi25a} for a schematic). Another key oxygen tracer is CO, as it will lock up the bulk of the free oxygen when C/O$>$1. Interestingly, CO -- at least the high $E_{\rm{up}}$ lines of the $P$-branch that is covered by MIRI MRS -- is not frequently detected in disks around VLMS: only two of the 10 VLMS sources in the MINDS program have CO in emission \citep{arabhavi25b}. It is unclear why oxygen is present in CO$_2$ but only weakly in H$_2$O or CO in these disks. Perhaps the H$_2$O and CO are present, but not readily observable, either due to the relative strength of the hydrocarbon $Q$-branches relative to the H$_2$O lines and/or photospheric absorption complicating the detection of disk CO emission (see \citealt{arabhavi25b}). 

Here we discuss some of the factors that may be driving this difference in chemistry between the very low-mass objects and the T Tauri sample.

\subsection{Stellar luminosity and the radiation field}\label{subsec: stellar radiation}

The stellar luminosity is the main heating source for protoplanetary disks. This energy input drives chemical evolution and equilibrium in the disks. The much lower luminosities of low-mass stars and brown dwarfs (\Lstar$\sim$0.015 to 0.1 \Lsun\ for the sample we explore here), mean that their disks are much colder and the snowlines are much closer to the central hosts. The flux of the H$_2$O lines follows the stellar luminosity, likely largely tracing a change in emitting area (\citealt{salyk11a} and Figure~\ref{fig: absolute fluxes}). On the other hand, the C$_2$H$_2$ flux does not follow this trend down to the very low-mass parameter space, indicating that the C$_2$H$_2$ flux is driven by something else besides emitting area (see Appendix D in \citealt{arabhavi25a} for further discussion).

Young low-mass stars and brown dwarfs will have different UV radiation spectra than Sun-like stars. However, young M dwarfs are also known to be very active, and X-rays and EUV from flares can have a significant impact on chemical evolution \citep{feinstein20}. All of these differences in the radiation impinging on the protoplanetary disk will impact the disk chemistry. 2D thermochemical modeling done by \cite{walsh15} finds that disks around M dwarfs are more molecule rich, compared to their higher-mass counterparts. This is due at least in part to the weaker far-UV radiation for these low-mass stars. Similarly, the models of \cite{woitke24} find that UV fluxes correlate with the line luminosities of OH and H$_2$O but there is a negative correlation with C$_2$H$_2$ and HCN due to photodissociation. Both modeling works find that X-ray-induced chemistry is important in setting the molecular complexity, in particular for C$_2$H$_2$ and HCN via the destruction of CO, N$_2$, and H$_2$. Therefore, the strong C$_2$H$_2$ emission in the VLMS disks may be at least partially caused by the weak UV fields of low-mass stars paired with their still moderate X-ray fluxes. \cite{sellek_vandishoeck25} explore the impact of ionization on the disk chemistry, in particular the destruction of CO by He$^{+}$, and find that the C liberated from CO can then go into the formation of hydrocarbons. This destruction pathway, paired with rapid radial transport, can result in high C/O in the inner disks of VLMS at relatively young ages. For this scenario to work, relatively high ionization rates are needed for the VLMS, potentially due to less shielding of the disks from cosmic rays by the stellar magnetic fields and/or higher ionization from the central stars.

UV emission generated by accretion of material from the disk onto the star will also be different between the T Tauris and VLMS. While the accretion mechanism is likely the same (i.e., magnetospheric accretion), the absolute accretion rate will decrease with decreasing stellar mass, and may decrease even more rapidly below \mstar$<$0.3 \Msun\ \citep{manara17}. \cite{colmenares24} argue that the carbon-rich disk around a solar-type star studied in their work may be due to a combination of carbon grain destruction paired with an unusually low accretion rate, which allows the carbon-rich gas to persist in the disk. As we show in Figure~\ref{fig: C2H2 H2O Mstar Mdot Mdust}, there is a relationship between $F_{\rm{C_2H_2}}$/$F_{\rm{H_2O}}$ and the accretion rate onto the star in our sample; however, while the degeneracies between the stellar mass and accretion rate make it difficult to determine what is truly driving the correlation, we do find that while $F_{\rm{H_2O}}$ decreases with decreasing accretion rate, the relationship between accretion rate and $F_{\rm{C_2H_2}}$ is not as strong (see Appendix A, Figure~\ref{fig: 10 mic Mdot Mdust separate}). Therefore, for the very low-mass objects, their low accretion rate may contribute to the high observed $F_{\rm{C_2H_2}}$/$F_{\rm{H_2O}}$ in two ways, by keeping the H$_2$O emitting area very small and by not removing all of the carbon-rich gas quickly. We note however that stars less than 0.3 \Msun\ have been found to have a steeper \LaccL\ relationship than higher mass sources, which may come from initially faster evolution \citep{manara17}. Thus, oxygen-rich gas may have been accreted fast at earlier ages and carbon-rich gas is accreting slowly by the age we are observing them (see discussion in \ref{subsubsec: dust drift}).

\subsection{Dust evolution}\label{subsec: dust evolution}

\subsubsection{Vertical dust settling}\label{subsubsec: settling}

Dust is the main opacity source in protoplanetary disks. At the earliest stages, the sub-micron sized particles will be held aloft in the disk atmosphere, but with time these grains grow from sub-micron particles into millimeter-sized grains to eventually pebbles, protoplanets, and the cores of giant planets. 

\textit{Spitzer}-IRS observations provided unique insights into the compositions and properties of dust in the inner disk for systems with a wide diversity of properties, including central hosts from the brown dwarf to Herbig Ae/Be regime. These works found that disks around low-mass stars and brown dwarfs show 10 \mic\ silicate emission features, which trace micron-sized dust grains in the disk atmosphere, that are weaker than their more massive counterparts and features that show fairly high levels of crystallinity \citep{apai05,kessler-silacci07,pascucci09}, now also found for objects with JWST \citep{kanwar24b,kaeufer24b}. However, the radial location in the disk where the 10 \mic\ feature will emit depends on the stellar luminosity, such that the emitting radius for brown dwarfs will be much closer to the central object than for Herbig Ae/Be stars ($\leq$0.001-0.1 au vs. $\geq$0.5-50 au; \citealt{kessler-silacci07}). Therefore, changes in the dust vertical scale, structure, or grain size distribution may just reflect differences in the probed emitting region, rather than differences due to the stellar influence or disk evolution. In addition to the 10 \mic\ silicate features being generally weaker in disks around VLMS, their strength may even be overestimated in some cases. Recent JWST observations have shown that broad features from 8.5 to 12.5 \mic\ can in fact come at least partially from high column densities of C$_2$H$_4$ \citep{arabhavi24a}, making it possible that some of the already weak silicate features seen in low-mass objects with \textit{Spitzer} could in fact not even be due to dust emission. However, when present, the VLMS dust features in our sample are similar in strength to T Tauri systems with weak features, supporting that we are seeing deeper into these disks. Further investigation of the dust properties in our VLMS sample will be included in separate work \citep{jang25_submitted}.

If the disks around low-mass hosts have undergone more efficient dust settling, such a lack of small dust grains in the disk atmosphere means that these disks will be subject to more stellar irradiation than those around their higher-mass counterparts, where the disk atmospheres are still relatively dust rich, which may lead to additional chemical reactions. However, the biggest implication of dust settling in low-mass sources is that mid-infrared observations probe more gas before hitting the $\tau_{MIR}$=1 surface.  This means that we are able to observe deeper into the disk than would otherwise be the case (see e.g., the case of J160532, \citealt{tabone23,franceschi24}, and the modeling work of \citealt{greenwood19,antonellini23,houge25a}). Therefore, we may be witnessing chemistry that may even be common in higher-mass systems, but which is usually hidden from view. If so, the optically thinner disks around low-mass sources may offer us a clearer picture of the midplane chemistry, where planet formation may be taking place. 2D thermochemical models find that the mid-infrared C$_2$H$_2$ emitting region is located deeper in the disk than H$_2$O (e.g., \citealt{woitke18b}), which, paired with the high columns of carbon-rich gas we are observing, supports the idea that we are probing closer to the disk midplane in disks around lower-mass objects (see Figure 9 in \citealt{arabhavi25b} for a schematic).

\subsubsection{Radial dust drift}\label{subsubsec: dust drift}

The inward drift of dust grains from the outer disk to the inner disk will impact the chemical composition in the inner disk. These dust grains begin their journey in the cold outer disk, coated in various molecular ices. As they travel inwards, they will pass snowlines of different molecular species, liberating their icy mantles as they go and enriching the gas along the way. This drift can then supply the inner disk with, for instance, an increase in H$_2$O vapor \citep{banzatti20,kalyaan21,banzatti23b}; however, it should be noted that this drift will also increase the amount of dust in the inner disk that can also shield the inner disk gas (e.g., \citealt{sellek24, houge25a}). \cite{banzatti20} found an anti-correlation between $F_{\rm{H_2O}}$/$F_{\rm{C_2H_2}}$ and $R_{\rm{dust}}$ (in our formulation with C$_2$H$_2$ in the numerator this would be a positive correlation: smaller disks have less C$_2$H$_2$ relative to H$_2$O). In their sample of T Tauri stars, this indicates that dust drift may be shrinking the dust disks and enriching the inner regions in H$_2$O. Conversely, if this drift is halted, for instance via gap opening by a giant planet, this can result in a relatively ``dry'' inner disk, that is lacking this extra oxygen enrichment \citep{najita13}.

Given this expectation that drift-dominated disks will be enriched in H$_2$O vapor, we might expect that disks around VLMS are rich in H$_2$O, due to their efficient dust drift (e.g., \citealt{pinilla13, pinilla22}). \cite{mah23} use disk evolution models to investigate this hypothesis and conclude that this enrichment might be so efficient and so rapid in low-mass systems, that this takes place very early in the disk lifetime. If this oxygen-enriched gas were to then accrete onto the central star/brown dwarf, in particular if accretion evolution happens faster in these sources \citep{manara17}, this would leave only carbon-rich gas to advect inward later and more slowly from the outer disk (e.g., \citealt{miotello19,bosman21}). This may result in a time-dependent change in the disk C/O ratio, where disks around low-mass hosts undergo a rapid decrease in C/O followed by a gradual increase in C/O (see also \citealt{sellek24}). This may also explain the 30 Myr evolved carbon-rich disk analyzed by \cite{long25}. The modeling work of \cite{sellek_vandishoeck25} find that disks that are compact initially, as may generally be the case for disks around VLMS, high C/O in the inner disk can be reached by young ages. However, if substructures were simply to slow down but not halt radial drift, this may prolong H$_2$O enrichment (e.g., \citealt{kalyaan23, mah24}), which is the proposed explanation for the H$_2$O-rich spectrum of a disk around a VLMS analyzed by \cite{xie23}. 
Although we cannot explore any correlations with dust radius for the VLMS sample due to the limited number of high angular resolution observations of these sources in general and in our sample particularly, we explore the relationship between $F_{\rm{C_2H_2}}$/$F_{\rm{H_2O}}$ and the disk dust mass (Figure~\ref{fig: C2H2 H2O Mstar Mdot Mdust}). As with the accretion rate, the trend here may be driven by the correlation between \mstar\ and \mdisk\ \citep{pascucci16}, and therefore with the stellar luminosity. However, the disk mass may in fact be a driving factor in setting the inner disk chemistry. More massive disks, which exist around more massive stars, will be able to form giant planets that are capable of forming traps that will halt radial drift (e.g., \citealt{vandermarel_mulders21}). The low-mass disks around low-mass stars, on the other hand, may be (generally) incapable of forming such deep substructures, therefore allowing radial drift to happen efficiently, bringing in ice-rich material quickly.

In order to observationally test the predicted connection between drift and inner disk chemistry in disks around very low-mass hosts, two steps would be paramount: 1) observations of disks around VLMS at very young ages ($\lesssim$0.5 Myr) to see if the gas is oxygen-rich and 2) high-resolution observations with ALMA to determine the outer radii and overall dust structure in disks around low-mass stars and brown dwarfs. The former can be done with JWST observations of younger star-forming regions, something that is currently being investigated (e.g., PID 3886, PI S. Grant), although going to even younger sources (i.e., embedded Class I sources) may be required (e.g., PID 7890, PI L. Tychoniec and PID  7135, PI K. Zhang). Exploring the latter -- the connection between outer disk substructures and inner disk chemistry in VLMS disks -- is challenging, given the faint nature of the disks around very low-mass stars and brown dwarfs. ALMA observations have only been done in a handful of sources at high enough angular resolution and sensitivity in the gas and dust to determine the $R_{\rm{gas}}$/$R_{\rm{dust}}$ ratio, which is expected to be high in the case of drift-dominated disks \citep{trapman19,toci21}. In shallow surveys these disks are often undetected, let alone spatially resolved. However, based on the small sample of resolved sources, it appears that drift is indeed quite efficient in disks around these objects \citep{kurtovic21}. That being said, there is generally a lack of overlap between the samples with JWST data and the high-resolution ALMA data to characterize the outer disks in these exceptionally faint objects. Efforts to correct this lack of overlap will be crucial to link outer disk processes to inner disk chemistry in these systems \citep{xie23} and some programs to do this are underway (e.g., 2024.1.00361.S, PI F. Long). Finally, while the focus of our discussion here has been on the early increase in C/O in low-mass systems, it should be noted that this scenario indicates that disks around T Tauri stars should show an increase in C/O at later times. Therefore, inner disk chemistry in T Tauri systems should be investigated across a range of ages to determine if such an increase in the C/O ratio is seen on timescales indicated by the models.

\subsubsection{Carbon-grain destruction}\label{subsubsec: carbon grain destruction}
Carbon-grain destruction is another viable method of increasing the gaseous C/O ratio in the inner disk (see the recent work of \citealt{houge25b}). While the sublimation temperature of refractory carbon species is debated in the literature, some estimates put the temperature as low as $\sim$500 K \citep{li21}. If this ``soot line'' were present in the disks around low-mass stars, the increase in gaseous carbon would allow for the formation of the carbon-rich gas species that are now observed \citep{tabone23,arabhavi24a}. While the average fit parameters of the hydrocarbons in the VLMS point to low temperatures (CO$_2$, HC$_3$N, and C$_6$H$_6$ all have temperatures below 250 K), thus potentially pointing away from this mechanism being the dominant factor in enhancing the carbon in these disks, it may instead be that the grains are not entirely destroyed and may instead simply be eroded by high-energy radiation or H atoms. If such destruction/erosion is happening in these disks, we may not see the same effects in most T Tauri disks either due to the higher dust opacities, blocking the C-rich gas from being observed, and/or due to the difference in the level of X-ray/cosmic ray ionization. While this may be generally true, the carbon-rich spectrum of DoAr 33 has been analyzed in this context by \cite{colmenares24}, who suggest that the low accretion rate of this object may allow the carbon-rich material to ``burn and linger'', which could also explain our VLMS sources, which have low accretion rates. However, we note that the correlation with accretion rate (Figure~\ref{fig: C2H2 H2O Mstar Mdot Mdust}) is not as strong as the correlation with stellar luminosity (see also Figure~\ref{fig: 10 mic Mdot Mdust separate}).

\section{Summary and conclusions}\label{sec: summary}

We explore the transition from H$_2$O-rich spectra to C$_2$H$_2$-dominated spectra with decreasing stellar luminosity using JWST-MIRI MRS observations from the MINDS collaboration, spanning (sub-)stellar masses from 0.02 to 1.5 \Msun.

\begin{enumerate}
    \item The flux of H$_2$O drops with decreasing stellar luminosity and accretion rate, as expected. Conversely, the C$_2$H$_2$ flux increases at the lowest host masses with an increase in the line-to-continuum ratio with decreasing \Lstar, driven by the strong C$_2$H$_2$ features present in the disks around low-mass stars and brown dwarfs. 
    \item We find a strong anti-correlation between the $F_{\rm{C_2H_2}}$/$F_{\rm{H_2O}}$ and stellar luminosity. Anti-correlations also exist between this flux ratio and the strength of the 10 \mic\ silicate feature, the accretion rate, and the disk mass, all of which may be due in part to strong correlations with the stellar mass and luminosity, but may be related to disk evolution processes that result in the abundant carbon-bearing species observed.
    \item We compute the average spectra for the very low-mass sample (VLMS, \mstar$\leq$0.2 \Msun) and the T Tauri sample (\mstar$>$0.2 \Msun). The average T Tauri spectrum is dominated by the forest of H$_2$O lines, but features from C$_2$H$_2$, HCN, CO$_2$, and OH are present. The average VLMS spectrum is dominated by the strong C$_2$H$_2$ $Q$-branch, with a molecular pseudo-continuum, with features from $^{13}$CCH$_2$, HCN, C$_6$H$_6$, CO$_2$, HC$_3$N, C$_3$H$_4$, C$_4$H$_2$, CH$_3$, and finally weak H$_2$O lines. 
    \item We use slab models to fit the average spectra for the T Tauri sample and the VLMS sample. We find that one component of the H$_2$O emission in the T Tauri average is similar in properties to the H$_2$O in the average VLMS spectrum, although the VLMS H$_2$O has a smaller emitting area, albeit in the optically thin regime where column density and emitting area are degenerate. Two components, one optically thick and one optically thin, are needed to reproduce the C$_2$H$_2$ emission in the VLMS spectrum, similar to what has been found in individual VLMS spectra analyzed thus far. The hydrocarbon gas is generally quite cold, with CH$_3$, HC$_3$N, C$_3$H$_4$, C$_4$H$_2$ and C$_6$H$_6$ having temperatures below $\sim$300 K.     
    \item We suggest that the $F_{\rm{C_2H_2}}$/$F_{\rm{H_2O}}$ correlation is driven by an increase in the volatile C/O ratio in the disks around very low-mass stars and brown dwarfs, although more modeling work is needed to explore exactly how $F_{\rm{C_2H_2}}$/$F_{\rm{H_2O}}$ and C/O are connected, especially in low-mass systems and utilizing extended carbon chemistry networks. If the high $F_{\rm{C_2H_2}}$/$F_{\rm{H_2O}}$ ratios for VLMS are due to a high gas phase C/O in the inner disk, this is likely driven by an enhancement in carbon-rich gas, rather than oxygen depletion alone. This carbon enrichment may be due to the weaker UV radiation and/or X-ray/cosmic ray ionization from very low-mass objects or due to the different evolutionary timescales of their disks, in particular the fast dust evolution that takes place in disks around VLMS, or some combination of multiple processes acting in concert. The latter can both alter the chemistry, via radial drift and/or by vertical dust settling, the latter of which allows for stellar radiation to penetrate closer to the disk midplane and means that our infrared observations probe deeper into the disk due to the decreased opacity. Some of these potential processes are tightly linked to the age of the systems, necessitating further exploration across a range of ages and evolutionary states in both stellar mass regimes. 

\end{enumerate}

By analyzing large samples of disks with a wide range of parameter space, we are able to investigate trends with system properties. However, given that many physical and chemical processes are correlated and, in some cases, interconnected, it is crucial to scrutinize the samples over multiple axes, to collect complementary data, and to compare the observations to models in order to determine the driving processes.

\begin{acknowledgements}
We thank the referee for constructive comments that improved the manuscript. This work is based on observations made with the NASA/ESA/CSA James Webb Space Telescope. The data were obtained from the Mikulski Archive for Space Telescopes at the Space Telescope Science Institute, which is operated by the Association of Universities for Research in Astronomy, Inc., under NASA contract NAS 5-03127 for JWST. These observations are associated with program \#1282. The following National and International Funding Agencies funded and supported the MIRI development: NASA; ESA; Belgian Science Policy Office (BELSPO); Centre Nationale d’Etudes Spatiales (CNES); Danish National Space Centre; Deutsches Zentrum fur Luft- und Raumfahrt (DLR); Enterprise Ireland; Ministerio De Econom\'ia y Competividad; Netherlands Research School for Astronomy (NOVA); Netherlands Organisation for Scientific Research (NWO); Science and Technology Facilities Council; Swiss Space Office; Swedish National Space Agency; and UK Space Agency.

E.v.D. acknowledges support from the ERC grant 101019751 MOLDISK and the Danish National Research Foundation through the Center of Excellence ``InterCat'' (DNRF150). M.T. and M.V. acknowledge support from the ERC grant 101019751 MOLDISK. A.M.A., I.K., and E.v.D. acknowledge support from grant TOP-1614.001.751 from the Dutch Research Council (NWO). B.T. is a Laureate of the Paris Region fellowship program (which is supported by the Ile-de-France Region) and has received funding under the Marie Sklodowska-Curie grant agreement No. 945298. T.H. and K.S. acknowledge support from the ERC Advanced Grant Origins 83 24 28. I.K. acknowledges funding from H2020-MSCA-ITN-2019, grant no. 860470 (CHAMELEON). A.C.G. has been supported by PRIN-INAF MAIN-STREAM 2017 and from PRIN-INAF 2019 (STRADE). V.C. acknowledges funding from the Belgian F.R.S.-FNRS. D.G. would like to thank the Research Foundation Flanders for co-financing the present research (grant number V435622N) and the European Space Agency (ESA) and the Belgian Federal Science Policy Office (BELSPO) for their support in the framework of the PRODEX Programme. T.K. acknowledges support from STFC Grant ST/Y002415/1. N.K. thanks the Deutsche Forschungsgemeinschaft (DFG) - grant 138 325594231, FOR 2634/2. M.M.C. has been funded by Spanish MCIN/AEI/10.13039/501100011033 grants PID2019-107061GB-C61 and No. MDM-2017-0737. G.P. gratefully acknowledges support from the Max Planck Society. L.M.S. has received funding from the European Research Council (ERC) under the European Union’s Horizon 2020 research and innovation programme (PROTOPLANETS, grant agreement No. 101002188).

\end{acknowledgements}

\bibliographystyle{aa}

\bibliography{aa55862-25}

\begin{thebibliography}{95}
\expandafter\ifx\csname natexlab\endcsname\relax\def\natexlab#1{#1}\fi

\bibitem[{{Alcal{\'a}} {et~al.}(2017){Alcal{\'a}}, {Manara}, {Natta}, {Frasca}, {Testi}, {Nisini}, {Stelzer}, {Williams}, {Antoniucci}, {Biazzo}, {Covino}, {Esposito}, {Getman}, \& {Rigliaco}}]{alcala17}
{Alcal{\'a}}, J.~M., {Manara}, C.~F., {Natta}, A., {et~al.} 2017, \aap, 600, A20

\bibitem[{{Anderson} {et~al.}(2021){Anderson}, {Blake}, {Cleeves}, {Bergin}, {Zhang}, {Schwarz}, {Salyk}, \& {Bosman}}]{anderson21}
{Anderson}, D.~E., {Blake}, G.~A., {Cleeves}, L.~I., {et~al.} 2021, \apj, 909, 55

\bibitem[{{Andrews} {et~al.}(2013){Andrews}, {Rosenfeld}, {Kraus}, \& {Wilner}}]{andrews13}
{Andrews}, S.~M., {Rosenfeld}, K.~A., {Kraus}, A.~L., \& {Wilner}, D.~J. 2013, \apj, 771, 129

\bibitem[{{Antonellini} {et~al.}(2023){Antonellini}, {Kamp}, \& {Waters}}]{antonellini23}
{Antonellini}, S., {Kamp}, I., \& {Waters}, L.~B.~F.~M. 2023, \aap, 672, A92

\bibitem[{{Apai} {et~al.}(2005){Apai}, {Pascucci}, {Bouwman}, {Natta}, {Henning}, \& {Dullemond}}]{apai05}
{Apai}, D., {Pascucci}, I., {Bouwman}, J., {et~al.} 2005, Science, 310, 834

\bibitem[{{Arabhavi} {et~al.}(2024){Arabhavi}, {Kamp}, {Henning}, {van Dishoeck}, {Christiaens}, {Gasman}, {Perrin}, {G{\"u}del}, {Tabone}, {Kanwar}, {Waters}, {Pascucci}, {Samland}, {Perotti}, {Bettoni}, {Grant}, {Lagage}, {Ray}, {Vandenbussche}, {Absil}, {Argyriou}, {Barrado}, {Boccaletti}, {Bouwman}, {Caratti o Garatti}, {Glauser}, {Lahuis}, {Mueller}, {Olofsson}, {Pantin}, {Scheithauer}, {Morales-Calder{\'o}n}, {Franceschi}, {Jang}, {Pawellek}, {Rodgers-Lee}, {Schreiber}, {Schwarz}, {Temmink}, {Vlasblom}, {Wright}, {Colina}, \& {{\"O}stlin}}]{arabhavi24a}
{Arabhavi}, A.~M., {Kamp}, I., {Henning}, T., {et~al.} 2024, Science, 384, 1086

\bibitem[{{Arabhavi} {et~al.}(2025{\natexlab{a}}){Arabhavi}, {Kamp}, {Henning}, {van Dishoeck}, {Jang}, {Waters}, {Christiaens}, {Gasman}, {Pascucci}, {Perotti}, {Grant}, {G{\"u}del}, {Lagage}, {Barrado}, {Garatti}, {Lahuis}, {Kaeufer}, {Kanwar}, {Morales-Calder{\'o}n}, {Schwarz}, {Sellek}, {Tabone}, {Temmink}, {Vlasblom}, \& {Patapis}}]{arabhavi25b}
{Arabhavi}, A.~M., {Kamp}, I., {Henning}, T., {et~al.} 2025{\natexlab{a}}, arXiv e-prints, arXiv:2506.02748

\bibitem[{{Arabhavi} {et~al.}(2025{\natexlab{b}}){Arabhavi}, {Kamp}, {van Dishoeck}, {Henning}, {Jang}, {Christiaens}, {Gasman}, {Pascucci}, {Perotti}, {Grant}, {Barrado}, {G{\"u}del}, {Lagage}, {Garatti}, {Lahuis}, {Waters}, {Kaeufer}, {Kanwar}, {Morales-Calder{\'o}n}, {Schwarz}, {Sellek}, {Tabone}, {Temmink}, \& {Vlasblom}}]{arabhavi25a}
{Arabhavi}, A.~M., {Kamp}, I., {van Dishoeck}, E.~F., {et~al.} 2025{\natexlab{b}}, arXiv e-prints, arXiv:2504.11425

\bibitem[{{Arulanantham} {et~al.}(2025){Arulanantham}, {Salyk}, {Pontoppidan}, {Banzatti}, {Zhang}, {{\"O}berg}, {Long}, {Carr}, {Najita}, {Pascucci}, {Jos{\'e} Colmenares}, {Xie}, {Huang}, {Green}, {Andrews}, {Blake}, {Bergin}, {Pinilla}, {Vioque}, {Dahl}, {Raul}, {Krijt}, \& {the JDISCS Collaboration}}]{arulanantham25}
{Arulanantham}, N., {Salyk}, C., {Pontoppidan}, K., {et~al.} 2025, arXiv e-prints, arXiv:2505.07562

\bibitem[{{Banzatti} {et~al.}(2020){Banzatti}, {Pascucci}, {Bosman}, {Pinilla}, {Salyk}, {Herczeg}, {Pontoppidan}, {Vazquez}, {Watkins}, {Krijt}, {Hendler}, \& {Long}}]{banzatti20}
{Banzatti}, A., {Pascucci}, I., {Bosman}, A.~D., {et~al.} 2020, \apj, 903, 124

\bibitem[{{Banzatti} {et~al.}(2023){Banzatti}, {Pontoppidan}, {Carr}, {Jellison}, {Pascucci}, {Najita}, {Mu{\~n}oz-Romero}, {{\"O}berg}, {Kalyaan}, {Pinilla}, {Krijt}, {Long}, {Lambrechts}, {Rosotti}, {Herczeg}, {Salyk}, {Zhang}, {Bergin}, {Ballering}, {Meyer}, {Bruderer}, \& {Jdiscs Collaboration}}]{banzatti23b}
{Banzatti}, A., {Pontoppidan}, K.~M., {Carr}, J.~S., {et~al.} 2023, \apjl, 957, L22

\bibitem[{{Banzatti} {et~al.}(2025){Banzatti}, {Salyk}, {Pontoppidan}, {Carr}, {Zhang}, {Arulanantham}, {Krijt}, {{\"O}berg}, {Cleeves}, {Najita}, {Pascucci}, {Blake}, {Romero-Mirza}, {Bergin}, {Cieza}, {Pinilla}, {Long}, {Mallaney}, {Xie}, {Waggoner}, {Kaeufer}, \& {The Jdiscs Collaboration}}]{banzatti25a}
{Banzatti}, A., {Salyk}, C., {Pontoppidan}, K.~M., {et~al.} 2025, \aj, 169, 165

\bibitem[{{Bosman} {et~al.}(2021){Bosman}, {Alarc{\'o}n}, {Bergin}, {Zhang}, {van't Hoff}, {{\"O}berg}, {Guzm{\'a}n}, {Walsh}, {Aikawa}, {Andrews}, {Bergner}, {Booth}, {Cataldi}, {Cleeves}, {Czekala}, {Furuya}, {Huang}, {Ilee}, {Law}, {Le Gal}, {Liu}, {Long}, {Loomis}, {M{\'e}nard}, {Nomura}, {Qi}, {Schwarz}, {Teague}, {Tsukagoshi}, {Yamato}, \& {Wilner}}]{bosman21}
{Bosman}, A.~D., {Alarc{\'o}n}, F., {Bergin}, E.~A., {et~al.} 2021, \apjs, 257, 7

\bibitem[{{Brown-Sevilla} {et~al.}(2021){Brown-Sevilla}, {Keppler}, {Barraza-Alfaro}, {Melon Fuksman}, {Kurtovic}, {Pinilla}, {Feldt}, {Brandner}, {Ginski}, {Henning}, {Klahr}, {Asensio-Torres}, {Cantalloube}, {Garufi}, {van Holstein}, {Langlois}, {M{\'e}nard}, {Rickman}, {Benisty}, {Chauvin}, {Zurlo}, {Weber}, {Pavlov}, {Ramos}, {Rochat}, \& {Roelfsema}}]{brown-sevilla21}
{Brown-Sevilla}, S.~B., {Keppler}, M., {Barraza-Alfaro}, M., {et~al.} 2021, \aap, 654, A35

\bibitem[{{Bushouse} {et~al.}(2024){Bushouse}, {Eisenhamer}, {Dencheva}, {Davies}, {Greenfield}, {Morrison}, {Hodge}, {Simon}, {Grumm}, {Droettboom}, {Slavich}, {Sosey}, {Pauly}, {Miller}, {Jedrzejewski}, {Hack}, {Davis}, {Crawford}, {Law}, {Gordon}, {Regan}, {Cara}, {MacDonald}, {Bradley}, {Shanahan}, {Jamieson}, {Teodoro}, {Williams}, {Pena-Guerrero}, {Graham}, {Molter}, {Brandt}, {Hayes}, {Cooper}, \& {Clarke}}]{bushouse_1.16.1}
{Bushouse}, H., {Eisenhamer}, J., {Dencheva}, N., {et~al.} 2024, {JWST Calibration Pipeline}

\bibitem[{{Carnall}(2017)}]{spectres}
{Carnall}, A.~C. 2017, arXiv e-prints, arXiv:1705.05165

\bibitem[{{Colmenares} {et~al.}(2024){Colmenares}, {Bergin}, {Salyk}, {Pontoppidan}, {Arulanantham}, {Calahan}, {Banzatti}, {Andrews}, {Blake}, {Ciesla}, {Green}, {Long}, {Lambrechts}, {Najita}, {Pascucci}, {Pinilla}, {Krijt}, {Trapman}, \& {Jdiscs Collaboration}}]{colmenares24}
{Colmenares}, M.~J., {Bergin}, E.~A., {Salyk}, C., {et~al.} 2024, \apj, 977, 173

\bibitem[{{Das} {et~al.}(2024){Das}, {Kurtovic}, \& {Flock}}]{das24}
{Das}, S., {Kurtovic}, N.~T., \& {Flock}, M. 2024, \aap, 689, A104

\bibitem[{{Erb}(2022)}]{erb22}
{Erb}, D. 2022, {pybaselines: A Python library of algorithms for the baseline correction of experimental data}

\bibitem[{{Facchini} {et~al.}(2021){Facchini}, {Teague}, {Bae}, {Benisty}, {Keppler}, \& {Isella}}]{facchini21}
{Facchini}, S., {Teague}, R., {Bae}, J., {et~al.} 2021, \aj, 162, 99

\bibitem[{{Fang} {et~al.}(2023){Fang}, {Pascucci}, {Edwards}, {Gorti}, {Hillenbrand}, \& {Carpenter}}]{fang23}
{Fang}, M., {Pascucci}, I., {Edwards}, S., {et~al.} 2023, \apj, 945, 112

\bibitem[{Feinstein {et~al.}(2020)Feinstein, Montet, Ansdell, Nord, Bean, Günther, Gully-Santiago, \& Schlieder}]{feinstein20}
Feinstein, A.~D., Montet, B.~T., Ansdell, M., {et~al.} 2020, The Astronomical Journal, 160, 219

\bibitem[{{Franceschi} {et~al.}(2024){Franceschi}, {Henning}, {Tabone}, {Perotti}, {Caratti o Garatti}, {Bettoni}, {van Dishoeck}, {Kamp}, {Absil}, {G{\"u}del}, {Olofsson}, {Waters}, {Arabhavi}, {Christiaens}, {Gasman}, {Grant}, {Jang}, {Rodgers-Lee}, {Samland}, {Schwarz}, {Temmink}, {Barrado}, {Boccaletti}, {Geers}, {Lagage}, {Pantin}, {Ray}, {Scheithauer}, {Vandenbussche}, \& {Wright}}]{franceschi24}
{Franceschi}, R., {Henning}, T., {Tabone}, B., {et~al.} 2024, \aap, 687, A96

\bibitem[{{Gangi} {et~al.}(2022){Gangi}, {Antoniucci}, {Biazzo}, {Frasca}, {Nisini}, {Alcal{\'a}}, {Giannini}, {Manara}, {Giunta}, {Harutyunyan}, {Munari}, \& {Vitali}}]{gangi22}
{Gangi}, M., {Antoniucci}, S., {Biazzo}, K., {et~al.} 2022, \aap, 667, A124

\bibitem[{{Gasman} {et~al.}(2023){Gasman}, {van Dishoeck}, {Grant}, {Temmink}, {Tabone}, {Henning}, {Kamp}, {G{\"u}del}, {Lagage}, {Perotti}, {Christiaens}, {Samland}, {Arabhavi}, {Argyriou}, {Abergel}, {Absil}, {Barrado}, {Boccaletti}, {Bouwman}, {Caratti o Garatti}, {Geers}, {Glauser}, {Guadarrama}, {Jang}, {Kanwar}, {Lahuis}, {Morales-Calder{\'o}n}, {Mueller}, {Nehm{\'e}}, {Olofsson}, {Pantin}, {Pawellek}, {Ray}, {Rodgers-Lee}, {Scheithauer}, {Schreiber}, {Schwarz}, {Vandenbussche}, {Vlasblom}, {Waters}, {Wright}, {Colina}, {Greve}, \& {{\"O}stlin}}]{gasman23b}
{Gasman}, D., {van Dishoeck}, E.~F., {Grant}, S.~L., {et~al.} 2023, \aap, 679, A117

\bibitem[{{Grant} {et~al.}(2022){Grant}, {Espaillat}, {Brittain}, {Scott-Joseph}, \& {Calvet}}]{grant22}
{Grant}, S.~L., {Espaillat}, C.~C., {Brittain}, S., {Scott-Joseph}, C., \& {Calvet}, N. 2022, \apj, 926, 229

\bibitem[{{Grant} {et~al.}(2024){Grant}, {Kurtovic}, {van Dishoeck}, {Henning}, {Kamp}, {Nowacki}, {Perraut}, {Banzatti}, {Temmink}, {Christiaens}, {Samland}, {Gasman}, {Tabone}, {G{\"u}del}, {Lagage}, {Arabhavi}, {Barrado}, {Caratti o Garatti}, {Glauser}, {Jang}, {Kanwar}, {Lahuis}, {Morales-Calder{\'o}n}, {Olofsson}, {Perotti}, {Schwarz}, {Vlasblom}, {Garcia Lopez}, \& {Long}}]{grant24b}
{Grant}, S.~L., {Kurtovic}, N.~T., {van Dishoeck}, E.~F., {et~al.} 2024, \aap, 689, A85

\bibitem[{{Grant} {et~al.}(2023){Grant}, {van Dishoeck}, {Tabone}, {Gasman}, {Henning}, {Kamp}, {G{\"u}del}, {Lagage}, {Bettoni}, {Perotti}, {Christiaens}, {Samland}, {Arabhavi}, {Argyriou}, {Abergel}, {Absil}, {Barrado}, {Boccaletti}, {Bouwman}, {o Garatti}, {Geers}, {Glauser}, {Guadarrama}, {Jang}, {Kanwar}, {Lahuis}, {Morales-Calder{\'o}n}, {Mueller}, {Nehm{\'e}}, {Olofsson}, {Pantin}, {Pawellek}, {Ray}, {Rodgers-Lee}, {Scheithauer}, {Schreiber}, {Schwarz}, {Temmink}, {Vandenbussche}, {Vlasblom}, {Waters}, {Wright}, {Colina}, {Greve}, {Justannont}, \& {{\"O}stlin}}]{grant23a}
{Grant}, S.~L., {van Dishoeck}, E.~F., {Tabone}, B., {et~al.} 2023, \apjl, 947, L6

\bibitem[{{Greenwood} {et~al.}(2019){Greenwood}, {Kamp}, {Waters}, {Woitke}, \& {Thi}}]{greenwood19}
{Greenwood}, A.~J., {Kamp}, I., {Waters}, L.~B.~F.~M., {Woitke}, P., \& {Thi}, W.~F. 2019, \aap, 626, A6

\bibitem[{{Haffert} {et~al.}(2019){Haffert}, {Bohn}, {de Boer}, {Snellen}, {Brinchmann}, {Girard}, {Keller}, \& {Bacon}}]{haffert19}
{Haffert}, S.~Y., {Bohn}, A.~J., {de Boer}, J., {et~al.} 2019, Nature Astronomy, 3, 749

\bibitem[{{Hartmann} {et~al.}(2016){Hartmann}, {Herczeg}, \& {Calvet}}]{hartmann16}
{Hartmann}, L., {Herczeg}, G., \& {Calvet}, N. 2016, \araa, 54, 135

\bibitem[{{Henning} {et~al.}(2024){Henning}, {Kamp}, {Samland}, {Arabhavi}, {Kanwar}, {van Dishoeck}, {Guedel}, {Lagage}, {Waelkens}, {Abergel}, {Absil}, {Barrado}, {Boccaletti}, {Bouwman}, {Garatti}, {Geers}, {Glauser}, {Lahuis}, {Nehme}, {Olofsson}, {Pantin}, {Ray}, {Vandenbussche}, {Waters}, {Wright}, {Christiaens}, {Franceschi}, {Gasman}, {Guadarrama}, {Jang}, {Morales-Calderon}, {Pawellek}, {Perotti}, {Rodgers-Lee}, {Schreiber}, {Schwarz}, {Tabone}, {Temmink}, {Vlasblom}, {Colina}, {Greve}, \& {Oestlin}}]{henning24}
{Henning}, T., {Kamp}, I., {Samland}, M., {et~al.} 2024, arXiv e-prints, arXiv:2403.09210

\bibitem[{{Herczeg} \& {Hillenbrand}(2014)}]{herczeg&hillenbrand14}
{Herczeg}, G.~J. \& {Hillenbrand}, L.~A. 2014, \apj, 786, 97

\bibitem[{{Hillenbrand} {et~al.}(1992){Hillenbrand}, {Strom}, {Vrba}, \& {Keene}}]{hillenbrand92}
{Hillenbrand}, L.~A., {Strom}, S.~E., {Vrba}, F.~J., \& {Keene}, J. 1992, \apj, 397, 613

\bibitem[{{Houge} {et~al.}(2025{\natexlab{a}}){Houge}, {Johansen}, {Bergin}, {Ciesla}, {Bitsch}, {Lambrechts}, {Henning}, \& {Perotti}}]{houge25b}
{Houge}, A., {Johansen}, A., {Bergin}, E., {et~al.} 2025{\natexlab{a}}, arXiv e-prints, arXiv:2505.20427

\bibitem[{{Houge} {et~al.}(2025{\natexlab{b}}){Houge}, {Krijt}, {Banzatti}, {Blake}, {Pinilla}, {Pontoppidan}, {Trapman}, {Williams}, \& {Zhang}}]{houge25a}
{Houge}, A., {Krijt}, S., {Banzatti}, A., {et~al.} 2025{\natexlab{b}}, \mnras, 537, 691

\bibitem[{{Jang} {et~al.}(2025){Jang}, {Arabhavi}, {Kaeufer}, {Waters}, Inga, {Henning}, {Caratti o Garatti}, {van Dishoeck}, {Perotti}, {Kanwar}, {Guedel}, {Morales-Calderón}, {Grant}, \& {Christiaens}}]{jang25_submitted}
{Jang}, H., {Arabhavi}, A.~M., {Kaeufer}, T., {et~al.} 2025, Submitted to \aap

\bibitem[{{Kaeufer} {et~al.}(2024{\natexlab{a}}){Kaeufer}, {Min}, {Woitke}, {Kamp}, \& {Arabhavi}}]{kaeufer24a}
{Kaeufer}, T., {Min}, M., {Woitke}, P., {Kamp}, I., \& {Arabhavi}, A.~M. 2024{\natexlab{a}}, \aap, 687, A209

\bibitem[{{Kaeufer} {et~al.}(2024{\natexlab{b}}){Kaeufer}, {Woitke}, {Kamp}, {Kanwar}, \& {Min}}]{kaeufer24b}
{Kaeufer}, T., {Woitke}, P., {Kamp}, I., {Kanwar}, J., \& {Min}, M. 2024{\natexlab{b}}, \aap, 690, A100

\bibitem[{{Kalyaan} {et~al.}(2023){Kalyaan}, {Pinilla}, {Krijt}, {Banzatti}, {Rosotti}, {Mulders}, {Lambrechts}, {Long}, \& {Herczeg}}]{kalyaan23}
{Kalyaan}, A., {Pinilla}, P., {Krijt}, S., {et~al.} 2023, \apj, 954, 66

\bibitem[{{Kalyaan} {et~al.}(2021){Kalyaan}, {Pinilla}, {Krijt}, {Mulders}, \& {Banzatti}}]{kalyaan21}
{Kalyaan}, A., {Pinilla}, P., {Krijt}, S., {Mulders}, G.~D., \& {Banzatti}, A. 2021, \apj, 921, 84

\bibitem[{{Kamp} {et~al.}(2023){Kamp}, {Henning}, {Arabhavi}, {Bettoni}, {Christiaens}, {Gasman}, {Grant}, {Morales-Calder{\'o}n}, {Tabone}, {Abergel}, {Absil}, {Argyriou}, {Barrado}, {Boccaletti}, {Bouwman}, {Caratti o Garatti}, {van Dishoeck}, {Geers}, {Glauser}, {G{\"u}del}, {Guadarrama}, {Jang}, {Kanwar}, {Lagage}, {Lahuis}, {Mueller}, {Nehm{\'e}}, {Olofsson}, {Pantin}, {Pawellek}, {Perotti}, {Ray}, {Rodgers-Lee}, {Samland}, {Scheithauer}, {Schreiber}, {Schwarz}, {Temmink}, {Vandenbussche}, {Vlasblom}, {Waelkens}, {Waters}, \& {Wright}}]{kamp23}
{Kamp}, I., {Henning}, T., {Arabhavi}, A.~M., {et~al.} 2023, Faraday Discussions, 245, 112

\bibitem[{{Kanwar} {et~al.}(2024){Kanwar}, {Kamp}, {Jang}, {Waters}, {van Dishoeck}, {Christiaens}, {Arabhavi}, {Henning}, {G{\"u}del}, {Woitke}, {Absil}, {Barrado}, {Caratti o Garatti}, {Glauser}, {Lahuis}, {Scheithauer}, {Vandenbussche}, {Gasman}, {Grant}, {Kurtovic}, {Perotti}, {Tabone}, \& {Temmink}}]{kanwar24b}
{Kanwar}, J., {Kamp}, I., {Jang}, H., {et~al.} 2024, \aap, 689, A231

\bibitem[{{Kessler-Silacci} {et~al.}(2006){Kessler-Silacci}, {Augereau}, {Dullemond}, {Geers}, {Lahuis}, {Evans}, {van Dishoeck}, {Blake}, {Boogert}, {Brown}, {J{\o}rgensen}, {Knez}, \& {Pontoppidan}}]{kessler-silacci06}
{Kessler-Silacci}, J., {Augereau}, J.-C., {Dullemond}, C.~P., {et~al.} 2006, \apj, 639, 275

\bibitem[{{Kessler-Silacci} {et~al.}(2007){Kessler-Silacci}, {Dullemond}, {Augereau}, {Mer{\'\i}n}, {Geers}, {van Dishoeck}, {Evans}, {Blake}, \& {Brown}}]{kessler-silacci07}
{Kessler-Silacci}, J.~E., {Dullemond}, C.~P., {Augereau}, J.~C., {et~al.} 2007, \apj, 659, 680

\bibitem[{{Kurtovic} {et~al.}(2021){Kurtovic}, {Pinilla}, {Long}, {Benisty}, {Manara}, {Natta}, {Pascucci}, {Ricci}, {Scholz}, \& {Testi}}]{kurtovic21}
{Kurtovic}, N.~T., {Pinilla}, P., {Long}, F., {et~al.} 2021, \aap, 645, A139

\bibitem[{{Li} {et~al.}(2021){Li}, {Bergin}, {Blake}, {Ciesla}, \& {Hirschmann}}]{li21}
{Li}, J., {Bergin}, E.~A., {Blake}, G.~A., {Ciesla}, F.~J., \& {Hirschmann}, M.~M. 2021, Science Advances, 7, eabd3632

\bibitem[{{Long} {et~al.}(2025){Long}, {Pascucci}, {Houge}, {Banzatti}, {Pontoppidan}, {Najita}, {Krijt}, {Xie}, {Williams}, {Herczeg}, {Andrews}, {Bergin}, {Blake}, {Colmenares}, {Harsono}, {Romero-Mirza}, {Li}, {Lu}, {Pinilla}, {Wilner}, {Vioque}, {Zhang}, \& {The Jdiscs Collaboration}}]{long25}
{Long}, F., {Pascucci}, I., {Houge}, A., {et~al.} 2025, \apjl, 978, L30

\bibitem[{{Luhman}(2007)}]{luhman07d}
{Luhman}, K.~L. 2007, \apjs, 173, 104

\bibitem[{{Mah} {et~al.}(2023){Mah}, {Bitsch}, {Pascucci}, \& {Henning}}]{mah23}
{Mah}, J., {Bitsch}, B., {Pascucci}, I., \& {Henning}, T. 2023, \aap, 677, L7

\bibitem[{{Mah} {et~al.}(2024){Mah}, {Savvidou}, \& {Bitsch}}]{mah24}
{Mah}, J., {Savvidou}, S., \& {Bitsch}, B. 2024, \aap, 686, L17

\bibitem[{{Manara} {et~al.}(2023){Manara}, {Ansdell}, {Rosotti}, {Hughes}, {Armitage}, {Lodato}, \& {Williams}}]{manara23}
{Manara}, C.~F., {Ansdell}, M., {Rosotti}, G.~P., {et~al.} 2023, in Astronomical Society of the Pacific Conference Series, Vol. 534, Protostars and Planets VII, ed. S.~{Inutsuka}, Y.~{Aikawa}, T.~{Muto}, K.~{Tomida}, \& M.~{Tamura}, 539

\bibitem[{{Manara} {et~al.}(2016){Manara}, {Rosotti}, {Testi}, {Natta}, {Alcal{\'a}}, {Williams}, {Ansdell}, {Miotello}, {van der Marel}, {Tazzari}, {Carpenter}, {Guidi}, {Mathews}, {Oliveira}, {Prusti}, \& {van Dishoeck}}]{manara16}
{Manara}, C.~F., {Rosotti}, G., {Testi}, L., {et~al.} 2016, \aap, 591, L3

\bibitem[{{Manara} {et~al.}(2017){Manara}, {Testi}, {Herczeg}, {Pascucci}, {Alcal{\'a}}, {Natta}, {Antoniucci}, {Fedele}, {Mulders}, {Henning}, {Mohanty}, {Prusti}, \& {Rigliaco}}]{manara17}
{Manara}, C.~F., {Testi}, L., {Herczeg}, G.~J., {et~al.} 2017, \aap, 604, A127

\bibitem[{{Manjavacas} {et~al.}(2024){Manjavacas}, {Tremblin}, {Birkmann}, {Valenti}, {Alves de Oliveira}, {Beck}, {Giardino}, {L{\"u}tzgendorf}, {Rauscher}, \& {Sirianni}}]{manjavacas24}
{Manjavacas}, E., {Tremblin}, P., {Birkmann}, S., {et~al.} 2024, \aj, 167, 168

\bibitem[{{Mendigut{\'\i}a} {et~al.}(2015){Mendigut{\'\i}a}, {Oudmaijer}, {Rigliaco}, {Fairlamb}, {Calvet}, {Muzerolle}, {Cunningham}, \& {Lumsden}}]{mendigutia15a}
{Mendigut{\'\i}a}, I., {Oudmaijer}, R.~D., {Rigliaco}, E., {et~al.} 2015, \mnras, 452, 2837

\bibitem[{{Min} {et~al.}(2011){Min}, {Dullemond}, {Kama}, \& {Dominik}}]{min11}
{Min}, M., {Dullemond}, C.~P., {Kama}, M., \& {Dominik}, C. 2011, \icarus, 212, 416

\bibitem[{{Miotello} {et~al.}(2019){Miotello}, {Facchini}, {van Dishoeck}, {Cazzoletti}, {Testi}, {Williams}, {Ansdell}, {van Terwisga}, \& {van der Marel}}]{miotello19}
{Miotello}, A., {Facchini}, S., {van Dishoeck}, E.~F., {et~al.} 2019, \aap, 631, A69

\bibitem[{{Morales-Calderón} {et~al.}(2025){Morales-Calderón}, {Jang}, {Arabhavi}, {Christiaens}, {Barrado}, {Kamp}, {van Dishoeck}, {Henning}, {Waters}, {Temmink}, {Guedel}, {Lagage}, {Caratti o Garatti}, {Glauser}, {Ray}, {Franceschi}, {Gasman}, {Grant}, {Kaeufer}, {Kanwar}, {Perotti}, {Samland}, {Schwarz}, {Vlasblom}, {Colina}, \& {Östlin}}]{morales-calderon25_submitted}
{Morales-Calderón}, M., {Jang}, H., {Arabhavi}, A.~M., {et~al.} 2025, In press in \aap

\bibitem[{{Mu{\~n}oz-Romero} {et~al.}(2024){Mu{\~n}oz-Romero}, {{\"O}berg}, {Banzatti}, {Pontoppidan}, {Andrews}, {Wilner}, {Bergin}, {Czekala}, {Law}, {Salyk}, {Teague}, {Qi}, {Bergner}, {Huang}, {Walsh}, {Guzm{\'a}n}, {Cleeves}, {Aikawa}, {Bae}, {Booth}, {Cataldi}, {Ilee}, {Le Gal}, {Long}, {Loomis}, {Menard}, \& {Liu}}]{munoz-romero24a}
{Mu{\~n}oz-Romero}, C.~E., {{\"O}berg}, K.~I., {Banzatti}, A., {et~al.} 2024, \apj, 964, 36

\bibitem[{{Mulders} {et~al.}(2015){Mulders}, {Ciesla}, {Min}, \& {Pascucci}}]{mulders15}
{Mulders}, G.~D., {Ciesla}, F.~J., {Min}, M., \& {Pascucci}, I. 2015, \apj, 807, 9

\bibitem[{{Najita} {et~al.}(2011){Najita}, {{\'A}d{\'a}mkovics}, \& {Glassgold}}]{najita11}
{Najita}, J.~R., {{\'A}d{\'a}mkovics}, M., \& {Glassgold}, A.~E. 2011, \apj, 743, 147

\bibitem[{{Najita} {et~al.}(2013){Najita}, {Carr}, {Pontoppidan}, {Salyk}, {van Dishoeck}, \& {Blake}}]{najita13}
{Najita}, J.~R., {Carr}, J.~S., {Pontoppidan}, K.~M., {et~al.} 2013, \apj, 766, 134

\bibitem[{{{\"O}berg} \& {Bergin}(2021)}]{oberg_bergin21}
{{\"O}berg}, K.~I. \& {Bergin}, E.~A. 2021, \physrep, 893, 1

\bibitem[{{Pascucci} {et~al.}(2009){Pascucci}, {Apai}, {Luhman}, {Henning}, {Bouwman}, {Meyer}, {Lahuis}, \& {Natta}}]{pascucci09}
{Pascucci}, I., {Apai}, D., {Luhman}, K., {et~al.} 2009, \apj, 696, 143

\bibitem[{{Pascucci} {et~al.}(2020){Pascucci}, {Banzatti}, {Gorti}, {Fang}, {Pontoppidan}, {Alexander}, {Ballabio}, {Edwards}, {Salyk}, {Sacco}, {Flaccomio}, {Blake}, {Carmona}, {Hall}, {Kamp}, {K{\"a}ufl}, {Meeus}, {Meyer}, {Pauly}, {Steendam}, \& {Sterzik}}]{pascucci20}
{Pascucci}, I., {Banzatti}, A., {Gorti}, U., {et~al.} 2020, \apj, 903, 78

\bibitem[{{Pascucci} {et~al.}(2013){Pascucci}, {Herczeg}, {Carr}, \& {Bruderer}}]{pascucci13}
{Pascucci}, I., {Herczeg}, G., {Carr}, J.~S., \& {Bruderer}, S. 2013, \apj, 779, 178

\bibitem[{{Pascucci} {et~al.}(2016){Pascucci}, {Testi}, {Herczeg}, {Long}, {Manara}, {Hendler}, {Mulders}, {Krijt}, {Ciesla}, {Henning}, {Mohanty}, {Drabek-Maunder}, {Apai}, {Sz{\H{u}}cs}, {Sacco}, \& {Olofsson}}]{pascucci16}
{Pascucci}, I., {Testi}, L., {Herczeg}, G.~J., {et~al.} 2016, \apj, 831, 125

\bibitem[{{Patapis} {et~al.}(2025){Patapis}, {Morales-Calder{\'o}n}, {Arabhavi}, {K{\"u}hnle}, {Gasman}, {Cugno}, {Molli{\`e} re}, {Matthews}, {M{\^a}lin}, {Whiteford}, {Lagage}, {Waters}, {Guedel}, {Henning}, {Vandenbussche}, {Absil}, {Argyriou}, {Barrado}, {Baudoz}, {Boccaletti}, {Bouwman}, {Cossou}, {Coulais}, {Decin}, {Gastaud}, {Glasse}, {Glauser}, {Grant}, {Min}, {Kamp}, {Olofsson}, {Pye}, {Rouan}, {Royer}, {Scheithauer}, {Sun}, {Tremblin}, {Colina}, {Ray}, {{\"O}stlin}, {van Dishoeck}, \& {Wright}}]{patapis25}
{Patapis}, P., {Morales-Calder{\'o}n}, M., {Arabhavi}, A.~M., {et~al.} 2025, arXiv e-prints, arXiv:2507.08961

\bibitem[{{Pecaut} \& {Mamajek}(2016)}]{pecaut_mamajek16}
{Pecaut}, M.~J. \& {Mamajek}, E.~E. 2016, \mnras, 461, 794

\bibitem[{{Perotti} {et~al.}(2023){Perotti}, {Christiaens}, {Henning}, {Tabone}, {Waters}, {Kamp}, {Olofsson}, {Grant}, {Gasman}, {Bouwman}, {Samland}, {Franceschi}, {van Dishoeck}, {Schwarz}, {G{\"u}del}, {Lagage}, {Ray}, {Vandenbussche}, {Abergel}, {Absil}, {Arabhavi}, {Argyriou}, {Barrado}, {Boccaletti}, {Caratti o Garatti}, {Geers}, {Glauser}, {Justannont}, {Lahuis}, {Mueller}, {Nehm{\'e}}, {Pantin}, {Scheithauer}, {Waelkens}, {Guadarrama}, {Jang}, {Kanwar}, {Morales-Calder{\'o}n}, {Pawellek}, {Rodgers-Lee}, {Schreiber}, {Colina}, {Greve}, {{\"O}stlin}, \& {Wright}}]{perotti23}
{Perotti}, G., {Christiaens}, V., {Henning}, T., {et~al.} 2023, \nat, 620, 516

\bibitem[{{Pinilla} {et~al.}(2013){Pinilla}, {Birnstiel}, {Benisty}, {Ricci}, {Natta}, {Dullemond}, {Dominik}, \& {Testi}}]{pinilla13}
{Pinilla}, P., {Birnstiel}, T., {Benisty}, M., {et~al.} 2013, \aap, 554, A95

\bibitem[{{Pinilla} {et~al.}(2022){Pinilla}, {Garufi}, \& {G{\'a}rate}}]{pinilla22}
{Pinilla}, P., {Garufi}, A., \& {G{\'a}rate}, M. 2022, arXiv e-prints, arXiv:2206.03057

\bibitem[{{Pontoppidan} {et~al.}(2024){Pontoppidan}, {Salyk}, {Banzatti}, {Zhang}, {Pascucci}, {{\"O}berg}, {Long}, {Mu{\~n}oz-Romero}, {Carr}, {Najita}, {Blake}, {Arulanantham}, {Andrews}, {Ballering}, {Bergin}, {Calahan}, {Cobb}, {Colmenares}, {Dickson-Vandervelde}, {Dignan}, {Green}, {Heretz}, {Herczeg}, {Kalyaan}, {Krijt}, {Pauly}, {Pinilla}, {Trapman}, \& {Xie}}]{pontoppidan24a}
{Pontoppidan}, K.~M., {Salyk}, C., {Banzatti}, A., {et~al.} 2024, \apj, 963, 158

\bibitem[{{Pontoppidan} {et~al.}(2014){Pontoppidan}, {Salyk}, {Bergin}, {Brittain}, {Marty}, {Mousis}, \& {{\"O}berg}}]{pontoppidan14a}
{Pontoppidan}, K.~M., {Salyk}, C., {Bergin}, E.~A., {et~al.} 2014, in Protostars and Planets VI, ed. H.~{Beuther}, R.~S. {Klessen}, C.~P. {Dullemond}, \& T.~{Henning}, 363

\bibitem[{{Romero-Mirza} {et~al.}(2024){Romero-Mirza}, {Banzatti}, {{\"O}berg}, {Pontoppidan}, {Salyk}, {Najita}, {Blake}, {Krijt}, {Arulanantham}, {Pinilla}, {Long}, {Rosotti}, {Andrews}, {Wilner}, {Calahan}, \& {The Jdiscs Collaboration}}]{romero-mirza24b}
{Romero-Mirza}, C.~E., {Banzatti}, A., {{\"O}berg}, K.~I., {et~al.} 2024, \apj, 975, 78

\bibitem[{{Salyk} {et~al.}(2025){Salyk}, {Pontoppidan}, {Banzatti}, {Bergin}, {Arulanantham}, {Najita}, {Blake}, {Carr}, {Zhang}, \& {Xie}}]{salyk25}
{Salyk}, C., {Pontoppidan}, K.~M., {Banzatti}, A., {et~al.} 2025, arXiv e-prints, arXiv:2502.05061

\bibitem[{{Salyk} {et~al.}(2011){Salyk}, {Pontoppidan}, {Blake}, {Najita}, \& {Carr}}]{salyk11a}
{Salyk}, C., {Pontoppidan}, K.~M., {Blake}, G.~A., {Najita}, J.~R., \& {Carr}, J.~S. 2011, \apj, 731, 130

\bibitem[{{Sellek} \& {van Dishoeck}(2025)}]{sellek_vandishoeck25}
{Sellek}, A.~D. \& {van Dishoeck}, E.~F. 2025, arXiv e-prints, arXiv:2507.11631

\bibitem[{{Sellek} {et~al.}(2024){Sellek}, {Vlasblom}, \& {van Dishoeck}}]{sellek24}
{Sellek}, A.~D., {Vlasblom}, M., \& {van Dishoeck}, E.~F. 2024, arXiv e-prints, arXiv:2412.01895

\bibitem[{{Tabone} {et~al.}(2023){Tabone}, {Bettoni}, {van Dishoeck}, {Arabhavi}, {Grant}, {Gasman}, {Henning}, {Kamp}, {G{\"u}del}, {Lagage}, {Ray}, {Vandenbussche}, {Abergel}, {Absil}, {Argyriou}, {Barrado}, {Boccaletti}, {Bouwman}, {Caratti o Garatti}, {Geers}, {Glauser}, {Justannont}, {Lahuis}, {Mueller}, {Nehm{\'e}}, {Olofsson}, {Pantin}, {Scheithauer}, {Waelkens}, {Waters}, {Black}, {Christiaens}, {Guadarrama}, {Morales-Calder{\'o}n}, {Jang}, {Kanwar}, {Pawellek}, {Perotti}, {Perrin}, {Rodgers-Lee}, {Samland}, {Schreiber}, {Schwarz}, {Colina}, {{\"O}stlin}, \& {Wright}}]{tabone23}
{Tabone}, B., {Bettoni}, G., {van Dishoeck}, E.~F., {et~al.} 2023, Nature Astronomy, 7, 805

\bibitem[{{Temmink} {et~al.}(2024{\natexlab{a}}){Temmink}, {van Dishoeck}, {Gasman}, {Grant}, {Tabone}, {G{\"u}del}, {Henning}, {Barrado}, {Caratti o Garatti}, {Glauser}, {Kamp}, {Arabhavi}, {Jang}, {Kurtovic}, {Perotti}, {Schwarz}, \& {Vlasblom}}]{temmink24b}
{Temmink}, M., {van Dishoeck}, E.~F., {Gasman}, D., {et~al.} 2024{\natexlab{a}}, \aap, 689, A330

\bibitem[{{Temmink} {et~al.}(2024{\natexlab{b}}){Temmink}, {van Dishoeck}, {Grant}, {Tabone}, {Gasman}, {Christiaens}, {Samland}, {Argyriou}, {Perotti}, {Guedel}, {Henning}, {Lagage}, {Abergel}, {Absil}, {Barrado}, {Garatti}, {Glauser}, {Kamp}, {Lahuis}, {Olofsson}, {Ray}, {Scheithauer}, {Vandenbussche}, {Waters}, {Arabhavi}, {Jang}, {Kanwar}, {Morales-Calderon}, {Rodgers-Lee}, {Schreiber}, {Schwarz}, \& {Colina}}]{temmink24a}
{Temmink}, M., {van Dishoeck}, E.~F., {Grant}, S.~L., {et~al.} 2024{\natexlab{b}}, arXiv e-prints, arXiv:2403.13591

\bibitem[{{Testi} {et~al.}(2022){Testi}, {Natta}, {Manara}, {Monsalvo}, {Lodato}, {Lopez}, {Muzic}, {Pascucci}, {Sanchis}, {Santamaria Miranda}, {Scholz}, {De Simone}, \& {Williams}}]{testi22}
{Testi}, L., {Natta}, A., {Manara}, C.~F., {et~al.} 2022, arXiv e-prints, arXiv:2201.04079

\bibitem[{{Toci} {et~al.}(2021){Toci}, {Rosotti}, {Lodato}, {Testi}, \& {Trapman}}]{toci21}
{Toci}, C., {Rosotti}, G., {Lodato}, G., {Testi}, L., \& {Trapman}, L. 2021, \mnras, 507, 818

\bibitem[{{Trapman} {et~al.}(2019){Trapman}, {Facchini}, {Hogerheijde}, {van Dishoeck}, \& {Bruderer}}]{trapman19}
{Trapman}, L., {Facchini}, S., {Hogerheijde}, M.~R., {van Dishoeck}, E.~F., \& {Bruderer}, S. 2019, \aap, 629, A79

\bibitem[{{van Boekel} {et~al.}(2003){van Boekel}, {Waters}, {Dominik}, {Bouwman}, {de Koter}, {Dullemond}, \& {Paresce}}]{vanboekel03}
{van Boekel}, R., {Waters}, L.~B.~F.~M., {Dominik}, C., {et~al.} 2003, \aap, 400, L21

\bibitem[{{van der Marel} \& {Mulders}(2021)}]{vandermarel_mulders21}
{van der Marel}, N. \& {Mulders}, G.~D. 2021, \aj, 162, 28

\bibitem[{{Vlasblom} {et~al.}(2025){Vlasblom}, {Temmink}, {Grant}, {Kurtovic}, {Sellek}, {van Dishoeck}, {G{\"u}del}, {Henning}, {Lagage}, {Barrado}, {Caratti o Garatti}, {Glauser}, {Kamp}, {Lahuis}, {Olofsson}, {Arabhavi}, {Christiaens}, {Gasman}, {Jang}, {Morales-Calder{\'o}n}, {Perotti}, {Schwarz}, \& {Tabone}}]{vlasblom25a}
{Vlasblom}, M., {Temmink}, M., {Grant}, S.~L., {et~al.} 2025, \aap, 693, A278

\bibitem[{{Walsh} {et~al.}(2015){Walsh}, {Nomura}, \& {van Dishoeck}}]{walsh15}
{Walsh}, C., {Nomura}, H., \& {van Dishoeck}, E. 2015, \aap, 582, A88

\bibitem[{{Wendeborn} {et~al.}(2024){Wendeborn}, {Espaillat}, {Lopez}, {Thanathibodee}, {Robinson}, {Pittman}, {Calvet}, {Flors}, {Walter}, {K{\'o}sp{\'a}l}, {Grankin}, {Mendigut{\'\i}a}, {G{\"u}nther}, {Eisl{\"o}ffel}, {Guo}, {France}, {Fiorellino}, {Fischer}, {{\'A}brah{\'a}m}, \& {Herczeg}}]{wendeborn24a}
{Wendeborn}, J., {Espaillat}, C.~C., {Lopez}, S., {et~al.} 2024, \apj, 970, 118

\bibitem[{{Woitke} {et~al.}(2018){Woitke}, {Min}, {Thi}, {Roberts}, {Carmona}, {Kamp}, {M{\'e}nard}, \& {Pinte}}]{woitke18b}
{Woitke}, P., {Min}, M., {Thi}, W.~F., {et~al.} 2018, \aap, 618, A57

\bibitem[{{Woitke} {et~al.}(2024){Woitke}, {Thi}, {Arabhavi}, {Kamp}, {K{\'o}sp{\'a}l}, \& {{\'A}brah{\'a}m}}]{woitke24}
{Woitke}, P., {Thi}, W.~F., {Arabhavi}, A.~M., {et~al.} 2024, \aap, 683, A219

\bibitem[{{Wright} {et~al.}(2023){Wright}, {Rieke}, {Glasse}, {Ressler}, {Garc{\'\i}a Mar{\'\i}n}, {Aguilar}, {Alberts}, {{\'A}lvarez-M{\'a}rquez}, {Argyriou}, {Banks}, {Baudoz}, {Boccaletti}, {Bouchet}, {Bouwman}, {Brandl}, {Breda}, {Bright}, {Cale}, {Colina}, {Cossou}, {Coulais}, {Cracraft}, {De Meester}, {Dicken}, {Engesser}, {Etxaluze}, {Fox}, {Friedman}, {Fu}, {Gasman}, {G{\'a}sp{\'a}r}, {Gastaud}, {Geers}, {Glauser}, {Gordon}, {Greene}, {Greve}, {Grundy}, {G{\"u}del}, {Guillard}, {Haderlein}, {Hashimoto}, {Henning}, {Hines}, {Holler}, {Detre}, {Jahromi}, {James}, {Jones}, {Justtanont}, {Kavanagh}, {Kendrew}, {Klaassen}, {Krause}, {Labiano}, {Lagage}, {Lambros}, {Larson}, {Law}, {Lee}, {Libralato}, {Lorenzo Alverez}, {Meixner}, {Morrison}, {Mueller}, {Murray}, {Mycroft}, {Myers}, {Nayak}, {Naylor}, {Nickson}, {Noriega-Crespo}, {{\"O}stlin}, {O'Sullivan}, {Ottens}, {Patapis}, {Penanen}, {Pietraszkiewicz}, {Ray}, {Regan}, {Roteliuk}, {Royer}, {Samara-Ratna}, {Samuelson}, {Sargent}, {Scheithauer},
  {Schneider}, {Schreiber}, {Shaughnessy}, {Sheehan}, {Shivaei}, {Sloan}, {Tamas}, {Teague}, {Temim}, {Tikkanen}, {Tustain}, {van Dishoeck}, {Vandenbussche}, {Weilert}, {Whitehouse}, \& {Wolff}}]{wright23}
{Wright}, G.~S., {Rieke}, G.~H., {Glasse}, A., {et~al.} 2023, \pasp, 135, 048003

\bibitem[{{Xie} {et~al.}(2023){Xie}, {Pascucci}, {Long}, {Pontoppidan}, {Banzatti}, {Kalyaan}, {Salyk}, {Liu}, {Najita}, {Pinilla}, {Arulanantham}, {Herczeg}, {Carr}, {Bergin}, {Ballering}, {Krijt}, {Blake}, {Zhang}, {{\"O}berg}, {Green}, \& {Jdiscs Collaboration}}]{xie23}
{Xie}, C., {Pascucci}, I., {Long}, F., {et~al.} 2023, \apjl, 959, L25

\end{thebibliography}

\clearpage

\onecolumn
\begin{appendix}

\section{$F_{\rm{C_2H_2}}$, $F_{\rm{H_2O}}$, and the line-to-continuum ratio}\label{sec: absolute fluxes}

\begin{figure*}[h]
    \centering
    \includegraphics[scale=0.57]{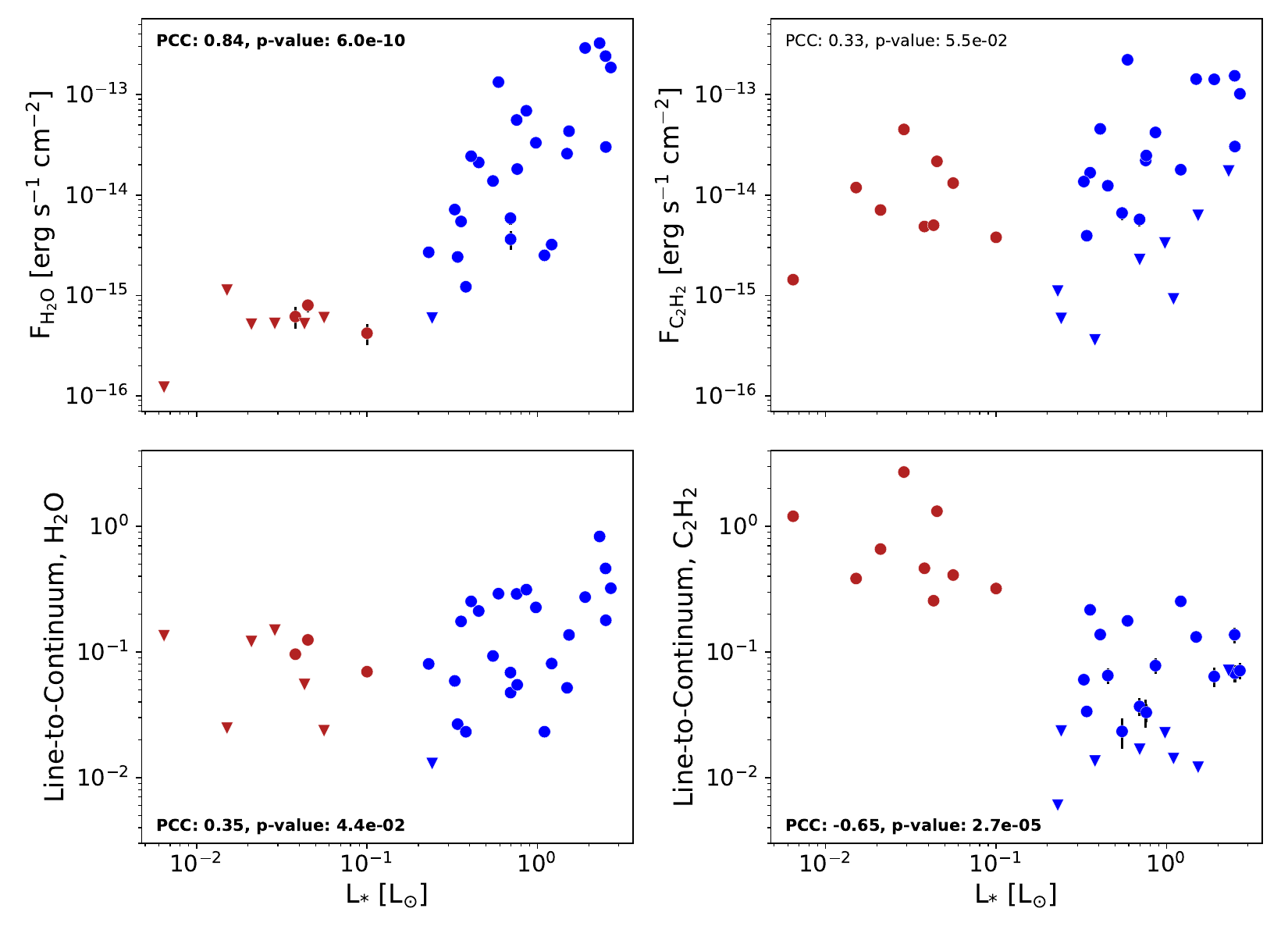}
    \caption{The absolute fluxes of H$_2$O (top left) and C$_2$H$_2$ (top right) as a function of stellar luminosity. The fluxes are determined after normalizing the spectra to a distance of 120 pc. The line-to-continuum ratio for H$_2$O and C$_2$H$_2$ as a function of stellar luminosity are shown in the bottom left and right, respectively. Upper limits (downward triangles) are the 3$\sigma$ fluxes. Error bars are smaller than the points for most targets. The blue points are the T Tauri sample and the red points are the VLMS targets. The PCCs and $p-$values can be found for each panel. Significant correlations ($p-$value$<$0.05) are provided in bold. $F_{\rm{H_2O}}$ vs. \Lstar\ (top left) and C$_2$H$_2$ line-to-continuum ratio vs. \Lstar\ (bottom right) have the strongest correlations. }
    \label{fig: absolute fluxes}
\end{figure*}

\begin{figure*}[h]
    \centering
    \includegraphics[scale=0.47]{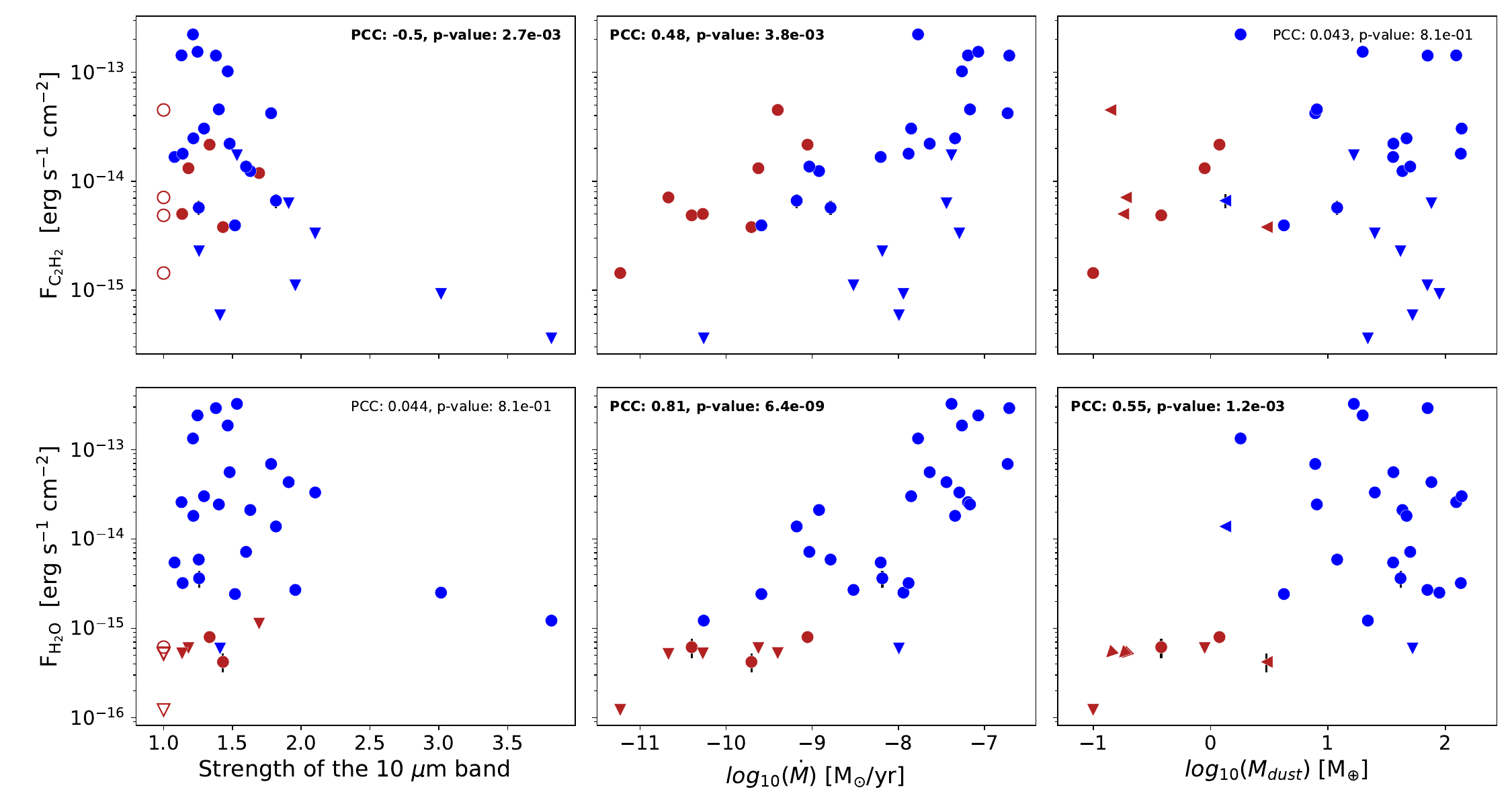}
    \caption{The absolute fluxes of C$_2$H$_2$ (top) and H$_2$O (bottom) as a function of the strength of the 10 $\mu$m silicate feature (left), stellar accretion rate (middle), and dust disk mass (right). The fluxes are normalized to a distance of 120 pc. Upper limits (downward triangles) are the 3$\sigma$ fluxes; objects with \Mdust\ upper limits are either leftward facing triangles if the molecular species is detected, or rotated triangular markers if both the flux and dust mass are upper limits. Error bars are smaller than the points for most targets. The blue points are the T Tauri sample and the red points are the VLMS targets. The PCCs and $p-$values can be found for each panel. Significant correlations ($p-$value$<$0.05) are provided in bold.}   
    \label{fig: 10 mic Mdot Mdust separate}
\end{figure*}

The relationships between the absolute flux values of C$_2$H$_2$ and H$_2$O and the line-to-continuum ratios for those molecules as a function of stellar luminosity are presented in Figure~\ref{fig: absolute fluxes}. The $F_{\rm{H_2O}}$--$L_*$ relationship is largely as expected: the lower luminosity of the host, the lower the line luminosity, although we may be limited by sensitivity for the H$_2$O lines in the VLMS sample. By contrast, there is no clear $F_{\rm{C_2H_2}}$--$L_*$ relationship; however, there is a correlation between the line-to-continuum ratio for C$_2$H$_2$ and the stellar luminosity. The relationships between the absolute fluxes and the strength of the 10 $\mu$m silicate feature, stellar accretion rate, and the disk dust mass are presented in Figure~\ref{fig: 10 mic Mdot Mdust separate}. The strongest correlation there is between the H$_2$O flux and the stellar accretion rate, found similarly between $L_{\rm{H_2O}}$ and the accretion luminosity by \cite{banzatti20}. However, the trend between the C$_2$H$_2$ line flux and the accretion rate, while present, is not very strong. This may be due to the bright C$_2$H$_2$ fluxes of the VLMS, as correlations have been seen between C$_2$H$_2$ and \Mdot\ in T Tauri samples by \cite{banzatti20} and \cite{arulanantham25}.

\cite{woitke24} find that the flux of C$_2$H$_2$ reacts more strongly to an increase in accretion rate than H$_2$O does, which is not what we find in our sample. Given the relationship between accretion rate and stellar luminosity (see e.g., \citealt{manara23} and references therein), we also check $F_{\rm{H_2O}}$ vs. \Mdot/\mstar, which are correlated, but with a weaker correlation than $F_{\rm{H_2O}}$--$\dot{M}$ and than $F_{\rm{C_2H_2}}$/$F_{\rm{H_2O}}$--$L_*$.

\section{$\chi^{2}$ maps}\label{sec: chi2 maps}
\begin{figure*}
    \centering
    \includegraphics[scale=0.55]{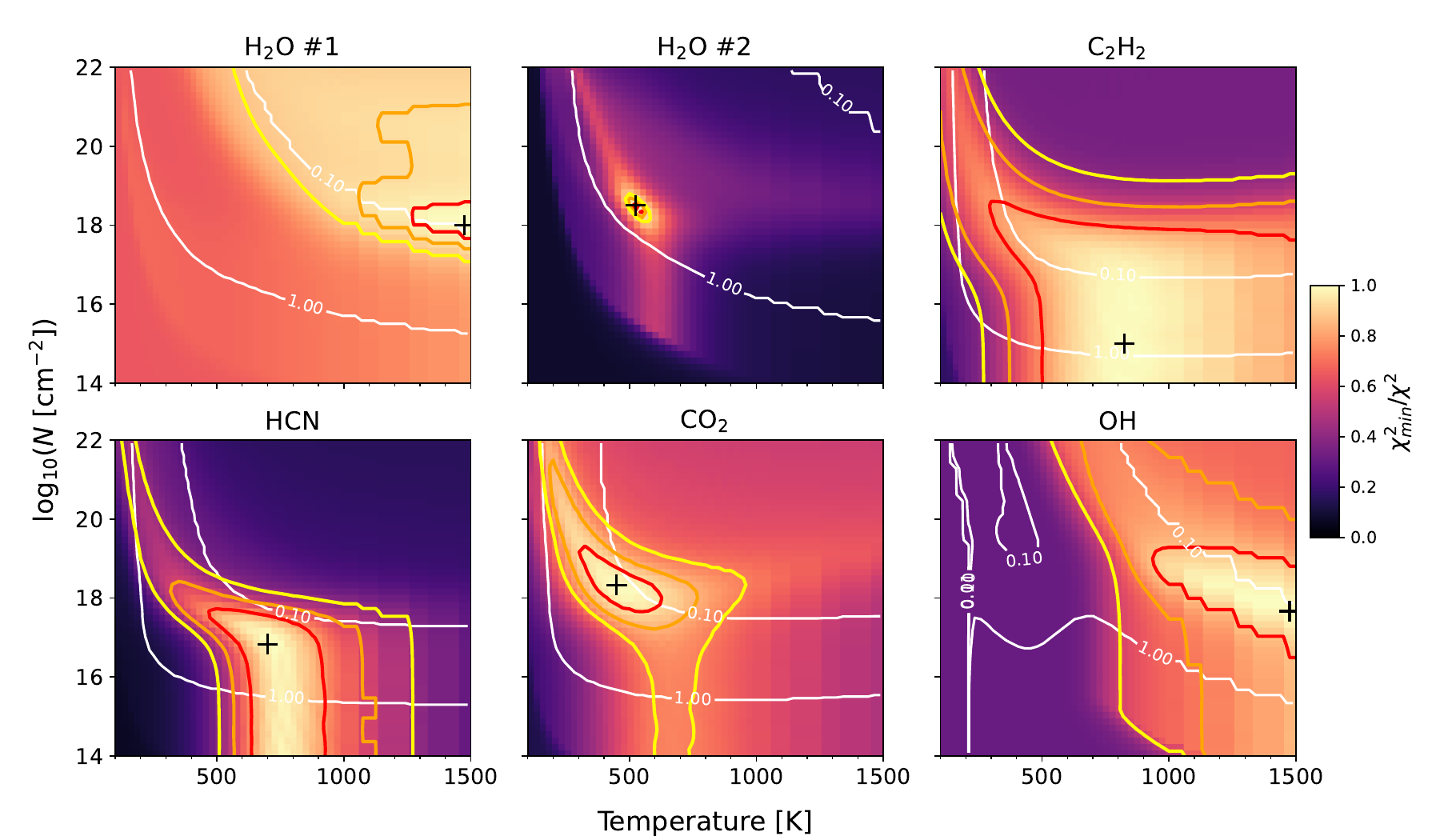}
    \caption{The $\chi^{2}$ maps for the molecules fitted in the average T Tauri spectrum. The color-scale shows $\chi^{2}_{min}/\chi^2$ and the red, orange, and yellow contours correspond to the 1$\sigma$, 2$\sigma$, and 3$\sigma$ levels. White contours show the emitting radii in au, as given by the labels. The best-fit model is represented by the black plus. }
    \label{fig: chi2 ttauri}
\end{figure*}

\begin{figure*}
    \centering
    \includegraphics[scale=0.55]{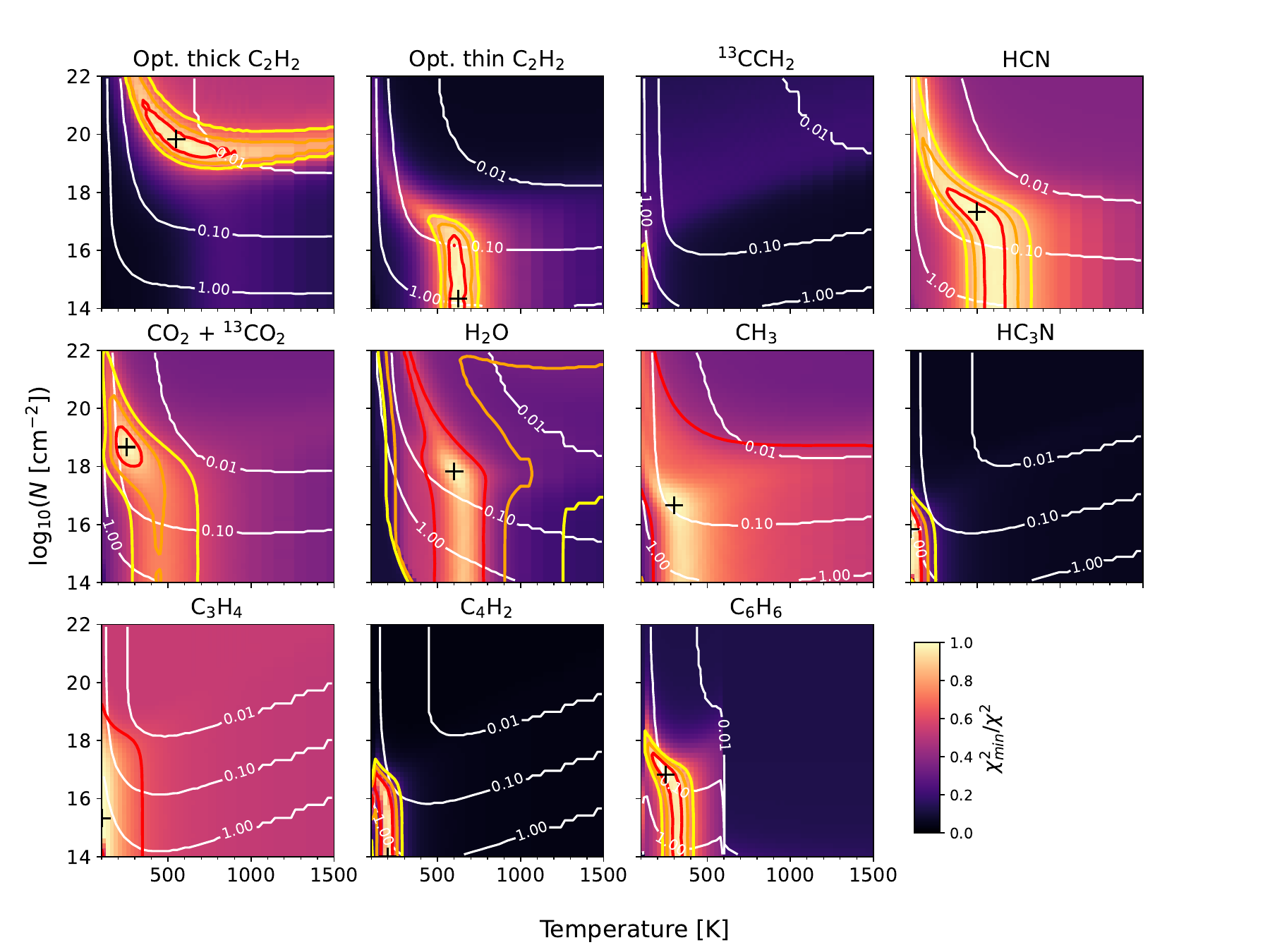}
    \caption{Same as Figure~\ref{fig: chi2 ttauri}, but now for the molecules in the average VLMS spectrum.}
    \label{fig: chi2 vlms}
\end{figure*}

The reduced $\chi^{2}$ maps for the molecules detected in the average T Tauri and VLMS spectra are shown in Figures~\ref{fig: chi2 ttauri} and \ref{fig: chi2 vlms}, respectively. The contours are the 1$\sigma$, 2$\sigma$, and 3$\sigma$ levels determined as $\chi^2_{min}$ + 2.3, $\chi^2_{min}$ + 6.2, and $\chi^2_{min}$ + 11.8, respectively (see Appendix C of \citealt{grant23a} for details). 

\end{appendix}

\end{document}